\documentclass[journal]{IEEEtran}
\usepackage{amsfonts,amssymb,amsmath,url,graphics,subfigure,cite,calc,psfrag,theorem}
\usepackage{amsmath}
\usepackage{here}
\usepackage{graphicx}
\usepackage{epstopdf}

    \makeatletter
\def\multilimits@{\bgroup
  \Let@
  \restore@math@cr
  \default@tag
 \baselineskip\fontdimen10 \scriptfont\tw@
 \advance\baselineskip\fontdimen12 \scriptfont\tw@
 \lineskip\thr@@\fontdimen8 \scriptfont\thr@@
 \lineskiplimit\lineskip
 \vbox\bgroup\ialign\bgroup\hfil$\m@th\scriptstyle{##}$\hfil\crcr}
\def\Sb{_\multilimits@}
\def\endSb{\crcr\egroup\egroup\egroup}
\makeatother
\newtheorem{theorem}{Theorem}

\newtheorem{definition}[theorem]{Definition}

\newtheorem{proposition}[theorem]{Proposition}

\input{tcilatex}
\begin{document}

\title{Labeled Random Finite Sets and the Bayes Multi-Target Tracking Filter
}
\author{\IEEEauthorblockN{Ba-Ngu Vo\IEEEauthorrefmark{1}, Ba-Tuong Vo\IEEEauthorrefmark{2}, Dinh Phung\IEEEauthorrefmark{3} }}

\maketitle

\begin{abstract}
An analytic solution to the multi-target Bayes recursion known as the $%
\delta $-Generalized Labeled Multi-Bernoulli ($\delta $-GLMB) filter has
been recently proposed in \cite{VoConj13}. As a sequel to \cite{VoConj13},
this paper details efficient implementations of the $\delta $-GLMB
multi-target tracking filter. Each iteration of this filter involves an
update operation and a prediction operation, both of which result in
weighted sums of multi-target exponentials with intractably large number of
terms. To truncate these sums, the ranked assignment and K-th shortest path
algorithms are used in the update and prediction, respectively, to determine
the most significant terms without exhaustively computing all of the terms.
In addition, using tools derived from the same framework, such as
probability hypothesis density filtering, we present inexpensive (relative
to the $\delta $-GLMB filter) look-ahead strategies to reduce the number of
computations. Characterization of the $L_{1}$-error in the multi-target
density arising from the truncation is presented.
\end{abstract}


\markboth{Preprint: IEEE Transactions on Signal Processing,~Vol.~62, No.~24, pp. 6554--6567,~2014}{Vo et al.: Labeled Random Finite Sets and the Bayes Multi-Target Tracking Filter}

%

~\renewcommand{\thefootnote}{} \footnotetext{%
\linespread{1.0} {\footnotesize
Acknowledgement: This work is supported by the Australian Research Council
under the Future Fellowship FT0991854 and Discovery Early Career Researcher Award DE120102388  \newline B.-N. Vo and B.-T. Vo are with the Department of Electrical and Computer Engineering, Curtin University,
Bentley, WA 6102, Australia (email: \url{{ba-ngu,ba-tuong}.vo@curtin.edu.au}). D. Phung is with the Faculty of Science, Engineering \& Built Environment, Deakin University, Geelong, VIC 3220, Australia (email: \url{dinh.phung@deakin.edu.au}) \newline}%
} \renewcommand{\thefootnote}{\arabic{footnote}}

\section{Introduction\label{sec:Intro}}

Multi-target filtering involves the on-line estimation of an unknown and
time-varying number of targets and their individual states from a sequence
of observations \cite{Blackman, Mahler07, Bar88}. While the term
multi-target filtering is often used interchangeably with multi-target
tracking, there is a subtle difference. In multi-target tracking we are
also interested in the trajectories of the targets (indeed, real
multi-target tracking systems require track labels). This work is concerned
with a Bayesian multi-target filtering solution that also provides estimates
of target trajectories, hence the name multi-target tracking filter.

The key challenges in multi-target filtering/tracking include \emph{%
detection uncertainty}, \emph{clutter}, and \emph{data association
uncertainty}. To date, three major approaches to multi-target
tracking/filtering have emerged as the main solution paradigms. These are,
Multiple Hypotheses Tracking (MHT), \cite{Reid77, Kurien90, Blackman,
Mallick12}, Joint Probabilistic Data Association (JPDA) \cite{Blackman,
Bar88}, and Random Finite Set (RFS) \cite{Mahler07}.

The random finite set (RFS) approach provides an elegant Bayesian
formulation of the multi-target filtering/tracking problem in which the
collection of target states, referred to as the multi-target state, is
treated as a finite set \cite{MahlerPHD2, Mahler07}. The rationale behind
this representation traces back to a fundamental consideration in estimation
theory--estimation error \cite{VVPS10}. This mathematical framework
subsequently became a very popular multi-target estimation method with
applications in sonar \cite{Clark07}, computer vision \cite{Pham07}, \cite%
{Maggio08}, \cite{Hosy12}, \cite{Hosy13}, field robotics \cite{Mullane11},
\cite{Zhang12}, \cite{Morat12}, \cite{Morat13}, \cite{Lee13} traffic
monitoring \cite{Battis08}, \cite{Mila13}, \cite{Meissner13}, cell biology
\cite{Juang09}, \cite{Hosy12}, \cite{Hamid13}, sensor network and
distributed estimation \cite{Zhang11}, \cite{LeeYao12}, \cite{Battis13},
\cite{Uney13} etc.

The centerpiece of the RFS approach is the \emph{Bayes multi-target filter}
\cite{Mahler07}, which recursively propagates the filtering density of the
multi-target state forward in time. This filter is also a (multi-target)
tracker when target identities or labels are incorporated into individual
target states. Due to the numerical complexity of Bayes multi-target filter,
the Probability Hypothesis Density (PHD) \cite{MahlerPHD2}, Cardinalized PHD
(CPHD) \cite{MahlerCPHDAES}, and multi-Bernoulli filters \cite{VVC09}, \cite%
{VVPS10} have been developed as approximations. These filters, in principle,
are not multi-target trackers because they rest on the premise that targets
are indistinguishable.

In \cite{VoVo11, VoConj13}, the notion of \emph{labeled RFSs} is introduced
to address target trajectories and their uniqueness. The key results include
conjugate priors that are closed under the Chapman-Kolmogorov equation, and
an analytic solution to the Bayes multi-target tracking filter known as the $%
\delta $-generalized labeled multi-Bernoulli ($\delta $-GLMB) filter.
Although a simulation result was presented to verify the solution, specific
implementation details were not given.

As a sequel to \cite{VoConj13}, the aim of this paper is to complement its
theoretical contributions with practical algorithms that will facilitate the
development of applications in signal processing and related fields. In
particular, we present an efficient and highly parallelizable implementation
of the $\delta $-GLMB filter. Each iteration of the $\delta $-GLMB filter
involves multi-target filtering and prediction densities that are weighted
sums of multi-target exponentials. While these sums are expressible in
closed forms, the number of terms grows super-exponentially in time.
Furthermore, it is not tractable to exhaustively compute all the terms of
the multi-target densities first and then truncate by discarding those
deemed insignificant.

The key innovation is the truncation of the multi-target densities without
exhaustively computing all their components. The multi-target filtering and
prediction densities are truncated using the ranked assignment and $K$%
-shortest paths algorithms, respectively. Techniques such as PHD
filtering are used as inexpensive look-ahead strategies to drastically
reduce the number of calls to ranked assignment and $K$-shortest paths algorithms. Moreover, we
establish that truncation by discarding $\delta $-GLMB components with small
weights minimizes the $L_{1}$ error in the multi-target density. To our best
knowledge, this is the first result regarding the effect of
truncation on the multi-target probability law.

The paper is organized as follows. Background on labeled RFS and the $\delta
$-GLMB filter is provided in section \ref{sec:BG}. Section \ref%
{sec:Truncationerror} establishes the $L_{1}$-distance between a $\delta $%
-GLMB density and its truncated version. Sections \ref{sec:delta-GLMB-MTT}
and \ref{sec:Prediction} present efficient implementations $\delta $-GLMB
filter update and prediction respectively. Section \ref{sec:Main} details
the multi-target state estimation, and discusses look-ahead strategies to
reduce the computational load. Numerical results are presented in Section %
\ref{sec:Simo} and concluding remarks are given in Section \ref%
{sec:Conclusion}.

\section{Background \label{sec:BG}}

This section summarizes the labeled RFS formulation of the multi-target
tracking problem and the $\delta $-GLMB filter proposed in \cite{VoConj13}.
Labeled RFS is summarized in subsection \ref{subsec:LabeledRFS}, followed by
Bayes multi-target filtering basics in subsections \ref{subsec:RFSBayes}, %
\ref{subsec:Likelihood} and \ref{subsec:Transition}. The $\delta $-GLMB
multi-target density and recursion are summarized in subsections \ref%
{subsec:delta-GLMB} and \ref{subsec:delta-GLMB filter}.

Throughout the paper, we use the standard inner product notation $%
\left\langle f,g\right\rangle \triangleq \int f(x)g(x)dx,$ and the following
multi-object exponential notation $h^{X}\triangleq \tprod\nolimits_{x\in
X}h(x)$, where $h$ is a real-valued function, with $h^{\emptyset }=1$ by
convention. We denote a generalization of the Kroneker delta that takes
arbitrary arguments such as sets, vectors, etc., by
\begin{equation*}
\delta _{Y}(X)\triangleq \left\{
\begin{array}{l}
1,\text{ if }X=Y \\
0,\text{ otherwise}%
\end{array}%
\right. ,
\end{equation*}%
and the inclusion function, a generalization of the indicator function, by%
\begin{equation*}
1_{Y}(X)\triangleq \left\{
\begin{array}{l}
1,\text{ if }X\subseteq Y \\
0,\text{ otherwise}%
\end{array}%
\right. .
\end{equation*}%
We also write $1_{Y}(x)$ in place of $1_{Y}(\{x\})$ when $X$ = $\{x\}$.

\subsection{Labeled RFS\label{subsec:LabeledRFS}}

An RFS is simply a finite-set-valued random variable \cite{Daley88}, \cite%
{Stoyanetal}. In this paper we use the Finite Set Statistics (FISST) notion
of integration/density to characterize RFSs \cite{MahlerPHD2, Mahler07}.
Treatments of RFS in the context of multi-target filtering can be found in
\cite{Mahler07, Vothesis08, Ristic13}.

To incorporate target identity, each state $x\in \mathbb{X}$ is augmented
with a unique label $\ell $ $\in $ $\mathbb{L}\mathcal{=\{\alpha }_{i}:i\in
\mathbb{N\}}$, where $\mathbb{N}$ denotes the set of positive integers and
the $\mathcal{\alpha }_{i}$'s are distinct.\

Let $\mathcal{L}:\mathbb{X}\mathcal{\times }\mathbb{L}\rightarrow \mathbb{L}$
be the projection $\mathcal{L}((x,\ell ))=\ell $, then a finite subset $%
\mathbf{X}$ of $\mathbb{X}\mathcal{\times }\mathbb{L}$ has distinct labels
if and only if $\mathbf{X}$ and its labels $\mathcal{L}(\mathbf{X})=\{%
\mathcal{L}(\mathbf{x})\!:\!\mathbf{x}\!\in \!\mathbf{X}\}$ have the same
cardinality, i.e. $\delta _{|\mathbf{X}|}(|\mathcal{L(}\mathbf{X})|)=1$. The
function $\Delta (\mathbf{X})\triangleq $ $\delta _{|\mathbf{X}|}(|\mathcal{%
L(}\mathbf{X})|)$ is called the \emph{distinct label indicator}.

\begin{definition}
A labeled RFS with state space $\mathbb{X}$ and (discrete) label space $%
\mathbb{L}$ is an RFS on $\mathbb{X}\mathcal{\times }\mathbb{L}$ such that
each realization has distinct labels.
\end{definition}

The unlabeled version of a labeled RFS is obtained by simply discarding the
labels. Consequently, the cardinality distribution (the distribution of the
number of objects) of a labeled RFS is the same as its unlabeled version.

For the rest of the paper, single-object states are represented by lowercase
letters (e.g. $x$, $\mathbf{x}$), while multi-object states are represented by
uppercase letters (e.g. $X$, $\mathbf{X}$), symbols for labeled states and
their distributions are bolded to distinguish them from unlabeled ones (e.g.
$\mathbf{x}$, $\mathbf{X}$, $\mathbf{\pi }$, etc.), spaces are represented by
blackboard bold (e.g. $\mathbb{X}$, $\mathbb{Z}$, $\mathbb{L}$, $\mathbb{N}$,
etc.), and the class of finite subsets of a space $\mathbb{X}$ is denoted by $%
\mathbf{\mathcal{F}(}\mathbb{X)}$. The integral of a function $\mathbf{f:}%
\mathbb{X}\mathcal{\times }\mathbb{L}\rightarrow \mathbb{R}$ is given by%
\begin{equation*}
\int \mathbf{f}(\mathbf{x})d\mathbf{x}=\sum_{\ell \in \mathbb{L}}\int_{%
\mathbb{X}}\mathbf{f}((x,\ell ))dx.
\end{equation*}

\subsection{Bayesian Multi-target Filtering\label{subsec:RFSBayes}}

\begin{figure}[tbh]
\begin{center}
\resizebox{80mm}{!}{\includegraphics[clip=true]{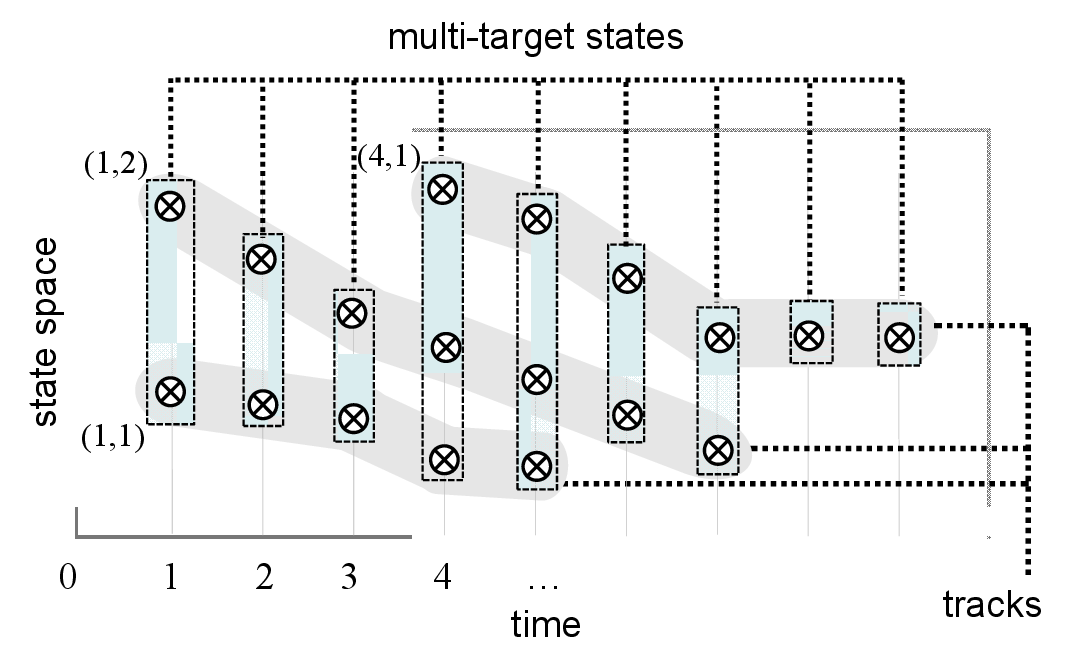}}
\end{center}
\caption{An example of label assignments. The two tracks born at time 1 are
given labels (1,1) and (1,2), while the only track born at time 4 is given
label (4,1). Notice also the difference between multi-target states and
target tracks.}
\label{fig:tracklabel}
\end{figure}

To incorporate target tracks in the Bayes multi-target filtering framework,
targets are identified by an ordered pair of integers $\ell =(k,i)$, where $%
k $ is the time of birth, and $i\in \mathbb{N}$ is a unique index to
distinguish objects born at the same time. Figure \ref{fig:tracklabel}
illustrates the assignment of labels to target trajectories. The label space
for objects born at time $k$, denoted as $\mathbb{L}_{k}$, is then $%
\{k\}\times \mathbb{N}$. An object born at time $k$, has state $\mathbf{x}%
\in \mathbb{X}\mathcal{\times }\mathbb{L}_{k}$. The label space for targets
at time $k$ (including those born prior to $k$), denoted as $\mathbb{L}%
_{0:k} $, is constructed recursively by $\mathbb{L}_{0:k}=\mathbb{L}%
_{0:k-1}\cup \mathbb{L}_{k}$. A multi-object state $\mathbf{X}$ at time $k$,
is a finite subset of $\mathbb{X}\mathcal{\times }\mathbb{L}_{0:k}$. Note
that $\mathbb{L}_{0:k-1}$ and $\mathbb{L}_{k}$ are disjoint.

Suppose that at time $k$, there are $N(k)$ target states $\mathbf{x}%
_{k,1},\ldots ,\mathbf{x}_{k,N(k)}$, each taking values in the (labeled)
state space $\mathbb{X\times L}$, and $M(k)$ measurements $z_{k,1},\ldots
,z_{k,M(k)}$ each taking values in an observation space $\mathbb{Z}$. In the
random finite set formulation, the set of targets and observations, at time $%
k$, \cite{MahlerPHD2, Mahler07} are treated as the \emph{multi-target state}
and \emph{multi-target observation}, respectively
\begin{align*}
\mathbf{X}_{k}& =\{\mathbf{x}_{k,1},\ldots ,\mathbf{x}_{k,N(k)}\}, \\
Z_{k}& =\{z_{k,1},\ldots ,z_{k,M(k)}\}.
\end{align*}

The \emph{multi-target posterior density} captures all information on the
set of target trajectories conditioned on the measurement history $%
Z_{0:k}=(Z_{0},...,Z_{k})$, and is given recursively for $k\geq 1$ by%
\begin{eqnarray*}
&&\!\!\!\!\!\!\!\!\!\!\!\!\!\!\!\!\!\!\!\!\!\mathbf{\pi }_{0:k}(\mathbf{X}%
_{0:k}|Z_{0:k}\mathbf{)}\text{ \ } \\
\text{ \ } &\propto &\!\!g_{k}(Z_{k}|\mathbf{X}_{k})\mathbf{f}_{k|k-1\!}(%
\mathbf{X}_{k}|\mathbf{X}_{k-1})\mathbf{\pi }_{0:k-1\!}(\mathbf{X}%
_{0:k-1}|Z_{0:k-1}\mathbf{),}
\end{eqnarray*}%
where $\mathbf{X}_{0:k}=(\mathbf{X}_{0},...,\mathbf{X}_{k})$, $g_{k}(\cdot
|\cdot )$ is the \emph{multi-target likelihood} \emph{function} at time $k$,
$\mathbf{f}_{k|k-1}(\cdot |\cdot )$ is the \emph{multi-target transition
density} to time $k$. The multi-target likelihood function encapsulates the
underlying models for detections and false alarms while the multi-target
transition density encapsulates the underlying models of target motions,
births and deaths.

Multi-target filtering is concerned with the marginal of the multi-target
posterior density, at the current time. Let $\mathbf{\pi }_{k}(\cdot |Z_{k})$
denote the \emph{multi-target filtering density }at time $k$\emph{, }and $%
\mathbf{\pi }_{k+1|k}$ denote the\emph{\ multi-target prediction density }to
time $k+1$ (formally $\mathbf{\pi }_{k}$ and $\mathbf{\pi }_{k+1|k}$ should
be written respectively as $\mathbf{\pi }_{k}(\cdot |Z_{0:k})$, and $\mathbf{%
\pi }_{k+1|k}(\cdot |Z_{0:k})$, but for simplicity we omit the dependence on
past measurements). Then, the \emph{multi-target Bayes filter} propagates $%
\mathbf{\pi }_{k}$ in time \cite{MahlerPHD2, Mahler07} according to the
following update and prediction
\begin{align}
\!\!\mathbf{\pi }_{k}(\mathbf{X}_{k}|Z_{k})& =\!\frac{g_{k}(Z_{k}|\mathbf{X}%
_{k})\mathbf{\pi }_{k|k-1}(\mathbf{X}_{k})}{\int g_{k}(Z_{k}|\mathbf{X})%
\mathbf{\pi }_{k|k-1}(\mathbf{X})\delta \mathbf{X}},  \label{eq:MTBayesPred}
\\
\!\!\mathbf{\pi }_{\!k+1|k\!}(\mathbf{X}_{k+1})& =\!\int \!\mathbf{f}%
_{\!k+1|k\!}(\mathbf{X}_{k+1}|\mathbf{X}_{k})\mathbf{\pi }_{k\!}(\mathbf{X}%
_{k}|Z_{k})\delta \!\mathbf{X}_{k},  \label{eq:MTBayesUpdate}
\end{align}%
where the integral is a \emph{set integral} defined for any function $%
\mathbf{f:\mathcal{F}(}\mathbb{X}\mathcal{\times }\mathbb{L)}\rightarrow
\mathbb{R}$ by%
\begin{equation*}
\int \mathbf{f}(\mathbf{X})\delta \mathbf{X}=\sum_{i=0}^{\infty }\frac{1}{i!}%
\int \mathbf{f}(\{\mathbf{x}_{1},...,\mathbf{x}_{i}\})d(\mathbf{x}_{1},...,%
\mathbf{x}_{i}).
\end{equation*}%
The multi-target filtering density captures all information on the
multi-target state, such as the number of targets and their states, at the
current time.

For convenience, in what follows we omit explicit references to the time
index $k$, and denote $\mathbb{L\triangleq L}_{0:k}$, $\mathbb{B\triangleq L}%
_{k+1}$, $\mathbb{L}_{+}\mathbb{\triangleq L}\cup \mathbb{B}$, $\mathbf{\pi
\mathbb{\triangleq }\pi }_{k}$, $\mathbf{\pi }_{+}\mathbb{\triangleq }%
\mathbf{\pi }_{k+1|k}$, $g\mathbb{\triangleq }g_{k},\mathbf{f}\mathbb{%
\triangleq }\mathbf{f}_{\!k+1|k\!}$.

\subsection{Measurement likelihood function\label{subsec:Likelihood}}

For a given multi-target state $\mathbf{X}$, at time $k$, each state $%
(x,\ell )\in \mathbf{X}$ is either detected with probability $p_{D}\left(
x,\ell \right) $ and generates a point $z$ with likelihood $g(z|x,\ell )$,
or missed with probability $1-p_{D}(x,\ell )$. The multi-object observation $%
Z=\{z_{1},...,z_{|Z|}\}$ is the superposition of the detected points and
Poisson clutter with intensity function $\kappa $.

\begin{definition}
An association map (for the current time) is a function $\theta :\mathbb{L}%
\rightarrow \{0,1,...,|Z|\}$ such that $\theta (i)=\theta (i^{\prime })>0~$%
implies$~i=i^{\prime }$. The set\ $\Theta $ of all such association maps is
called the association map space. The subset of association maps with domain
$I$ is denoted by $\Theta (I)$.
\end{definition}

An association map describes which tracks generated which measurements, i.e.
track $\ell $ generates measurement $z_{\theta (\ell )}\in Z$, with
undetected tracks assigned to $0$. The condition $\theta (i)=\theta
(i^{\prime })>0~$implies$~i=i^{\prime }$, means that a track can generate at
most one measurement at any point in time.

Assuming that, conditional on $\mathbf{X}$, detections are independent, and
that clutter is independent of the detections, the multi-object likelihood
is given by%
\begin{equation}
g(Z|\mathbf{X})=e^{-\left\langle \kappa ,1\right\rangle }\kappa
^{Z}\sum_{\theta \in \Theta (\mathcal{L(}\mathbf{X}))}\left[ \psi
_{\!Z}(\cdot ;\theta )\right] ^{\mathbf{X}}  \label{eq:RFSmeaslikelihood0}
\end{equation}%
where
\begin{equation}
\psi _{Z}(x,\ell ;\theta )=\left\{
\begin{array}{ll}
\frac{p_{D}(x,\ell )g(z_{\theta (\ell )}|x,\ell )}{\kappa (z_{\theta (\ell
)})}, & \text{if }\theta (\ell )>0 \\
1-p_{D}(x,\ell ), & \text{if }\theta (\ell )=0%
\end{array}%
\right.  \label{eq:PropConj5}
\end{equation}%
Equation (\ref{eq:RFSmeaslikelihood0}) is equivalent to the likelihood
function given by (54) in \cite{VoConj13}, and is more convenient for
implementation.

\subsection{Multi-target transition kernel\label{subsec:Transition}}

Given the current multi-object state $\mathbf{X}$,\ each state $(x,\ell )$ $%
\in \mathbf{X}$ either continues to exist at the next time step with
probability $p_{S}(x,\ell )$ and evolves to a new state $(x_{+},\ell _{+})$
with probability density $f(x_{+}|x,\ell )\delta _{\ell }(\ell _{+})$, or
dies\ with probability $1-p_{S}(x,\ell )$. The set of new targets born at
the next time step is distributed according to%
\begin{equation}
\mathbf{f}_{B}(\mathbf{Y})=\Delta (\mathbf{Y})w_{B}(\mathcal{L(}\mathbf{Y}))%
\left[ p_{B}\right] ^{\mathbf{Y}}  \label{eq:Birth_transition}
\end{equation}%
where $w_{B}$ and $p_{B}$\ are given parameters of the multi-target birth
density $\mathbf{f}_{B}$, defined on $\mathbb{X}\times \mathbb{B}$. Note
that $\mathbf{f}_{B}(\mathbf{Y})=0$ if $\mathbf{Y}$ contains any element $%
\mathbf{y}$ with $\mathcal{L(}\mathbf{y})\notin \mathbb{B}$. The birth model
(\ref{eq:Birth_transition}) covers both labeled Poisson and labeled
multi-Bernoulli \cite{VoConj13}.

The multi-target state at the next time $\mathbf{X}_{+}$ is the
superposition of surviving targets and new born targets. Assuming that
targets evolve independently of each other and that births are independent
of surviving targets, it was shown in \cite{VoConj13} that the multi-target
transition kernel is given by
\begin{equation}
\mathbf{f}\left( \mathbf{X}_{+}|\mathbf{X}\right) =\mathbf{f}_{S}(\mathbf{X}%
_{+}\cap (\mathbb{X}\times \mathbb{L)}|\mathbf{X})\mathbf{f}_{B}(\mathbf{X}%
_{+}-(\mathbb{X}\times \mathbb{L}))  \label{eq:labeled_transition}
\end{equation}%
where
\begin{eqnarray}
\!\!\mathbf{f}_{S}(\mathbf{W}|\mathbf{X})\!\!\! &=&\!\!\!\Delta (\mathbf{W}%
)\Delta (\mathbf{X})1_{\mathcal{L}(\mathbf{X})}(\mathcal{L(}\mathbf{W}))%
\left[ \Phi (\mathbf{W};\cdot )\right] ^{\mathbf{X}}
\label{eq:Survival_transition} \\
\!\!\Phi (\mathbf{W};x,\ell )\!\!\! &=&\!\!\!\left\{ \!\!%
\begin{array}{ll}
p_{S}(x,\ell )f(x_{+}|x,\ell ), & \!\!\text{if }\left( x_{+},\ell \right)
\in \mathbf{W} \\
1-p_{S}(x,\ell ), & \!\!\text{if }\ell \notin \mathcal{L}(\mathbf{W})%
\end{array}%
\right. \!\!.  \label{eq:Survival_transition2}
\end{eqnarray}

\subsection{Delta-Generalized Labeled Multi-Bernoulli \label%
{subsec:delta-GLMB}}

The $\delta $-generalized labeled multi-Bernoulli filter is a solution to
the Bayes multi-target filter based on the family of generalized labeled
multi-Bernoulli (GLMB) distributions%
\begin{equation*}
\mathbf{\pi }(\mathbf{X})=\Delta (\mathbf{X})\sum_{\xi \in \Xi }w^{(\xi )}(%
\mathcal{L(}\mathbf{X}))\left[ p^{(\xi )}\right] ^{\mathbf{X}},
\end{equation*}%
where $\Xi $ is a discrete space, each $p^{(\xi )}(\cdot ,\ell )$ is a
probability density, and each $w^{(\xi )}(I)$ is non-negative with $%
\sum_{(I,\xi )\in \mathcal{F}\!(\mathbb{L})\!\times \!\Xi }w^{(\xi )}(I)=1$.
A GLMB can be interpreted as a mixture of multi-target exponentials \cite%
{VoConj13}. While this family is closed under the Bayes recursion \cite%
{VoConj13}, it is not clear how numerical implementation can be
accomplished. Fortunately, an alternative form of the GLMB, known as $\delta
$-GLMB%
\begin{equation}
\mathbf{\pi }(\mathbf{X})=\Delta (\mathbf{X})\sum_{(I,\xi )\in \mathcal{F}(%
\mathbb{L})\times \Xi }\omega ^{(I,\xi )}\delta _{I}(\mathcal{L(}\mathbf{X}))%
\left[ p^{(\xi )}\right] ^{\mathbf{X}},  \label{eq:generativeGLMB}
\end{equation}%
where $\omega ^{(I,\xi )}=w^{(\xi )}(I)$, provides a representation that
facilitates numerical implementation. Note that the $\delta $-GLMB can be
obtained from the GLMB by using the identity $w^{(\xi )}(J)=\sum_{I\in
\mathcal{F}\!(\mathbb{L})}w^{(\xi )}(I)\delta _{I}(J)$, since the summand is
non-zero if and only if $I=J$.

In the $\delta $-GLMB initial multi-target prior
\begin{equation}
\mathbf{\pi }_{0}(\mathbf{X})=\Delta (\mathbf{X})\sum_{I\in \mathcal{F}(%
\mathbb{L}_{0})}\omega _{0}^{(I)}\delta _{I}(\mathcal{L(}\mathbf{X}))p_{0}^{%
\mathbf{X}},  \label{eq:deltaGLMBinitial}
\end{equation}%
each $I\in \mathcal{F}(\mathbb{L}_{0})$ represents a set of tracks labels
born at time $0$, $\omega _{0}^{(I)}$ represents the weight of the
hypothesis that $I$ is the set of track labels at time $0$, and $p_{0}(\cdot
,\ell )$ is the probability density of the kinematic state of track $\ell
\in I$. For example, suppose that there are 2 possibilities:

\begin{enumerate}
\item 0.3 chance of 1 target with label $(0,2)$, and density $p_{0}(\cdot
,(0,2))=\mathcal{N}(\cdot ;m,P_{2})$,

\item 0.7 chance of 2 targets with labels $(0,1)$, $(0,2)$ and respective
densities $p_{0}(\cdot ,(0,1))=\mathcal{N}(\cdot ;0,P_{1})$, $p_{0}(\cdot
,(0,2))=\mathcal{N}(\cdot ;m,P_{2})$.
\end{enumerate}

Then the $\delta $-GLMB representation is%
\begin{equation*}
\!\mathbf{\pi }_{0\!}(\mathbf{X})\!=\!0.3\delta _{\{(0,2)\!\}}(\mathcal{L(}%
\mathbf{X}))p_{0}^{\mathbf{X}}+0.7\delta _{\{(0,1),(0,2)\!\}}(\mathcal{L(}%
\mathbf{X}))p_{0}^{\mathbf{X}}\!.
\end{equation*}%
Note that the initial prior (\ref{eq:deltaGLMBinitial}) is a $\delta $-GLMB
with $\Xi =\emptyset $. For $\delta $-GLMB filtering and prediction
densities that are conditioned on measurements up to time $k$, the discrete
space $\Xi $ is the space of association map histories $\Theta
_{0:k}\triangleq \Theta _{0}\times ...\times \Theta _{k}$, where $\Theta _{t}
$ denotes the association map space at time $t$. In particular, as shown in
\cite{VoConj13}, for each $k\geq 0$ the filtering and prediction densities
are $\delta $-GLMB densities:%
\begin{eqnarray}
\!\!\!\!\!\!\mathbf{\pi }_{\!k\!}(\mathbf{X|}Z_{k\!})\!\!\!\!\!
&=&\!\!\!\!\!\Delta \!(\mathbf{X})\!\!\!\!\!\!\!\!\!\!\sum_{(I,\xi )\in
\mathcal{F}(\mathbb{L}_{0:k})\times \Theta
_{0:k}}\!\!\!\!\!\!\!\!\!\!\!\!\omega _{k}^{(I,\xi )}\delta _{\!I}(\mathcal{%
L(}\mathbf{X}))\!\left[ p_{k}^{(\xi )}(\cdot \mathbf{|}Z_{k\!})\right] ^{\!%
\mathbf{X}}  \label{eq:deltaGLMBposterior} \\
\!\!\!\!\!\!\mathbf{\pi }_{\!k\!+\!1|k\!}(\mathbf{X})\!\!\!\!\!
&=&\!\!\!\!\!\Delta \!(\mathbf{X})\!\!\!\!\!\!\!\!\!\!\!\!\!\sum_{(I,\xi
)\in \mathcal{F}(\mathbb{L}_{0:k+1})\times \Theta
_{0:k}}\!\!\!\!\!\!\!\!\!\!\!\!\omega _{k+1|k}^{(I,\xi )}\delta _{\!I}(%
\mathcal{L(}\mathbf{X}))\!\left[ p_{k\!+\!1|k}^{(\xi )}\right] ^{\!\mathbf{X}%
}  \label{eq:deltaGLMBprediction}
\end{eqnarray}%
Each $I\in \mathcal{F}(\mathbb{L}_{0:k})$ represents a set of track labels
at time $k$, and each $\xi =(\theta _{0},...,\theta _{k})\in \Theta _{0:k}$
represents a history of association maps up to time $k$, which also contains
the history of target labels encapsulating both births and deaths. The pair $%
(I,\xi )\in \mathcal{F}(\mathbb{L}_{0:k})\times \Theta _{0:k}$ is called a
\emph{hypothesis}, and its associated weight $\omega _{k}^{(I,\xi )}$ can be
interpreted as the probability of the hypothesis. Similarly the pair $(I,\xi
)\in \mathcal{F}(\mathbb{L}_{0:k+1})\times \Theta _{0:k}$ is called a \emph{%
prediction hypothesis}, with probability $\omega _{k+1|k}^{(I,\xi )}$. The
densities $p_{k}^{(\xi )}(\cdot ,\ell )$ and $p_{k\!+\!1|k}^{(\xi )}(\cdot
,\ell )$ are the filtering and prediction densities of the kinematic state
of track $\ell $ for association map history $\xi $.

\subsection{Delta Generalized Labeled Multi-Bernoulli Recursion\label%
{subsec:delta-GLMB filter}}

The $\delta $-GLMB filter recursively propagates a $\delta $-GLMB filtering
density forward in time via the Bayes update and prediction equations (\ref%
{eq:MTBayesUpdate}) and (\ref{eq:MTBayesPred}). Closed form solutions to the
update and prediction of the $\delta $-GLMB filter are given by the
following results \cite{VoConj13}.

\begin{proposition}
\label{PropBayes_strong}If the current multi-target prediction density is a $%
\delta $-GLMB of the form (\ref{eq:generativeGLMB}), then the multi-target
filtering density is a $\delta $-GLMB given by%
\begin{equation}
\mathbf{\pi }\!(\mathbf{X|}Z)=\Delta \!(\mathbf{X})\!\!\!\!\!\!\!\!\sum_{(I,%
\xi )\in \mathcal{F}\!(\mathbb{L})\!\times \!\Xi }\;\sum\limits_{\theta
\!\in \Theta \!(I)}\!\!\!\!\omega ^{\!(I,\xi ,\theta \!)\!}(Z)\delta
_{\!I\!}(\mathcal{L\!(}\mathbf{X})\!)\!\!\left[ p^{\!(\xi ,\theta )\!}(\cdot
|Z)\right] ^{\!\mathbf{X}}  \label{eq:PropBayes_strong0}
\end{equation}%
where $\Theta (I)$ denotes the subset of current association maps with
domain $I$,\allowdisplaybreaks%
\begin{eqnarray}
\omega ^{(I,\xi ,\theta )\!}(Z)\!\!\! &\propto &\!\!\!\omega ^{(I,\xi
)}[\eta _{Z}^{(\xi ,\theta )}]^{I},  \label{eq:PropBayes_strong1} \\
\eta _{Z}^{(\xi ,\theta )}(\ell )\!\!\! &=&\!\!\!\left\langle p^{(\xi
)}(\cdot ,\ell ),\psi _{Z}(\cdot ,\ell ;\theta )\right\rangle ,
\label{eq:PropBayes_strong2} \\
p^{\!(\xi ,\theta )\!}(x,\ell |Z)\!\!\! &=&\!\!\!\frac{p^{(\xi )}(x,\ell
)\psi _{Z}(x,\ell ;\theta )}{\eta _{Z}^{(\xi ,\theta )}(\ell )},
\label{eq:PropBayes_strong3}
\end{eqnarray}
\end{proposition}

\begin{proposition}
\label{Prop_CK_strong}If the current multi-target filtering density is a $%
\delta $-GLMB of the form (\ref{eq:generativeGLMB}), then the multi-target
prediction to the next time is a $\delta $-GLMB given by%
\begin{equation}
\mathbf{\pi }_{\!+}(\mathbf{X}_{\!+\!})=\Delta \!(\mathbf{X}%
_{\!+})\!\!\!\!\!\!\sum_{(I_{+},\xi )\in \mathcal{F}(\mathbb{L}_{+})\times
\Xi }\!\!\omega _{+}^{(I_{+},\xi )}\delta _{I_{+\!}}(\mathcal{L(}\mathbf{X}%
_{\!+}))\!\left[ p_{+}^{(\xi )\!}\right] ^{\!\mathbf{X}_{+}}
\label{eq:PropCKstrong1}
\end{equation}%
where\allowdisplaybreaks%
\begin{eqnarray}
\!\!\!\omega _{+}^{(I_{+},\xi )}\!\! &=&\!\!\omega _{S}^{(\xi )}(I_{+}\cap
\mathbb{L})w_{B}(I_{+}\cap \mathbb{B})  \label{eq:PropCKstrong2} \\
\!\!\!\omega _{S}^{(\xi )}(L)\!\! &=&\!\![\eta _{S}^{(\xi
)}]^{L}\sum_{I\supseteq L}[1-\eta _{S}^{(\xi )}]^{I-L}\omega ^{(I,\xi )}
\label{eq:PropCKstrongws} \\
\!\!\!\eta _{S}^{(\xi )}(\ell )\!\! &=&\!\!\left\langle p_{S}(\cdot ,\ell
),p^{(\xi )}(\cdot ,\ell )\right\rangle   \label{eq:PropCKstrong_eta} \\
\!\!\!p_{+}^{(\xi )}(x,\ell )\!\! &=&\!\!1_{\mathbb{L}}(\ell )p_{S}^{(\xi
)\!}(x,\ell )+1_{\mathbb{B}\!}(\ell )p_{B}(x,\ell )  \label{eq:PropCKstrong3}
\\
\!\!\!p_{S}^{(\xi )}(x,\ell )\!\! &=&\!\!\frac{\left\langle p_{S}(\cdot
,\ell )f(x|\cdot ,\ell ),p^{(\xi )}(\cdot ,\ell )\right\rangle }{\eta
_{S}^{(\xi )}(\ell )}  \label{eq:PropCKstrong4}
\end{eqnarray}
\end{proposition}

Note from Propositions \ref{PropBayes_strong} and \ref{Prop_CK_strong} that
the actual value of the association history $\xi $ is not used in the
calculations, it is used merely as an indexing variable. On the other hand,
the value of the label set $I$ is used in the calculations.

\section{Characterizing Truncation Error\label{sec:Truncationerror}}

\begin{figure}[tbh]
\begin{center}
\resizebox{80mm}{!}{\includegraphics[clip=true]{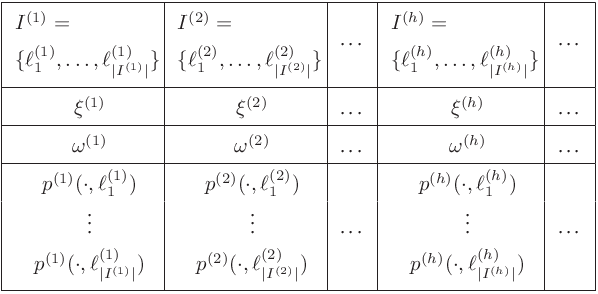}}
\end{center}
\caption{An enumeration of a $\protect\delta $-GLMB parameter set with each
component indexed by an integer $h$. The hypothesis\ for component $h$ is $%
(I^{(h)},\protect\xi ^{(h)})$ while its weight and associated track
densities are $\protect\omega ^{(h)}$ and $p^{(h)}(\cdot ,\ell )$, $\ell \in
I^{(h)}$.}
\label{fig:paramtable}
\end{figure}

A $\delta $-GLMB is completely characterized by the set of parameters $%
\{(\omega ^{(I,\xi )},p^{(\xi )}):(I,\xi )\in \mathcal{F}\!(\mathbb{L}%
)\!\times \!\Xi \}$. For implementation it is convenient to consider the set
of $\delta $-GLMB parameters as an enumeration of all hypotheses (with
positive weight) together with their associated weights and track densities $%
\{(I^{(h)},\xi ^{(h)},\omega ^{(h)},p^{(h)})\}_{h=1}^{H}$, as shown in
Figure \ref{fig:paramtable}, where $\omega ^{(h)}\triangleq \omega
^{(I^{(h)},\xi ^{(h)})}$ and $p^{(h)}\triangleq p^{(\xi ^{(h)})}$.
Implementing the $\delta $-GLMB filter then amounts to recursively
propagating the set of $\delta $-GLMB parameters forward in time.

Since the number of hypotheses grows super-exponentially with time, it is
necessary to reduce the number of components in the $\delta $-GLMB parameter
set, at each time step. A simple solution is to truncate the $\delta $-GLMB
density by discarding ``insignificant'' hypotheses.

The effect of truncation on the probability law of the multi-target state
has not been characterized even though this truncation is widely used in MHT
and JPDA. In the RFS approach the probability law of the multi-target state
is completely captured in the multi-target density. Consequently, the effect
of discarding hypotheses in a $\delta $-GLMB density can be characterized by
the difference/dissimilarity between the untruncated and truncated $\delta $%
-GLMB densities. The following result establishes the $L_{1}$-error between
a $\delta $-GLMB density and its truncated version.

\begin{proposition}
\label{Prop_L1_error}Let $\left\Vert \mathbf{f}\right\Vert _{1}\triangleq
\int \left\vert \mathbf{f}(\mathbf{X})\right\vert \delta \mathbf{X}$ denote
the $L_{1}$-norm of $\mathbf{f:\mathcal{F}(}\mathbb{X}\mathcal{\times }%
\mathbb{L)}\rightarrow \mathbb{R}$, and for a given $\mathbb{H\subseteq }%
\mathcal{F}(\mathbb{L})\times \Xi $ let
\begin{equation*}
\mathbf{f}_{\mathbb{H}}\mathbf{(X)}=\Delta (\mathbf{X})\sum\limits_{(I,\xi
)\in \mathbb{H}}\omega ^{(I,\xi )}\delta _{I}(\mathcal{L(}\mathbf{X}))\left[
p^{(\xi )}\right] ^{\mathbf{X}}
\end{equation*}%
be an unnormalized $\delta $-GLMB density (i.e. does not necessarily
integrate to 1). If $\mathbb{T\subseteq H}$ then%
\begin{eqnarray*}
||\mathbf{f}_{\mathbb{H}}-\mathbf{f}_{\mathbb{T}}||_{1} &\mathbf{=}%
&\sum\limits_{(I,\xi )\in \mathbb{H-T}}\omega ^{(I,\xi )}, \\
\left\Vert \frac{\mathbf{f}_{\mathbb{H}}}{||\mathbf{f}_{\mathbb{H}}||_{1}}-%
\frac{\mathbf{f}_{\mathbb{T}}}{||\mathbf{f}_{\mathbb{T}}\mathbf{||}_{1}}%
\right\Vert _{1} &\leq &2\frac{||\mathbf{f}_{\mathbb{H}}||_{1}-||\mathbf{f}_{%
\mathbb{T}}\mathbf{||}_{1}}{||\mathbf{f}_{\mathbb{H}}||_{1}},
\end{eqnarray*}
\end{proposition}

Proof:\allowdisplaybreaks%
\begin{eqnarray*}
&&\!\!\!\!\!\!\!\!\!\!\!\!\!\!\!\!\!\!\left\Vert \mathbf{f}_{\mathbb{H}}-%
\mathbf{f}_{\mathbb{T}}\right\Vert _{1} \\
&=&\int \left\vert \Delta (\mathbf{X})\sum\limits_{(I,\xi )\in \mathbb{H-T}%
}\omega ^{(I,\xi )}\delta _{I}(\mathcal{L(}\mathbf{X}))\left[ p^{(\xi )}%
\right] ^{\mathbf{X}}\right\vert \delta \mathbf{X} \\
&=&\sum\limits_{(I,\xi )\in \mathbb{H-T}}\!\!\!\omega ^{(I,\xi )}\int \Delta
(\mathbf{X})\delta _{I}(\mathcal{L(}\mathbf{X}))\left[ p^{(\xi )}\right] ^{%
\mathbf{X}}\delta \mathbf{X}\text{ \ \ } \\
&=&\sum\limits_{(I,\xi )\in \mathbb{H-T}}\!\!\!\omega ^{(I,\xi
)}\sum_{L\subseteq \mathbb{L}}\delta _{I}(L),\text{ \ \ \ (by Lemma 3 of
\cite{VoConj13})} \\
&=&\sum\limits_{(I,\xi )\in \mathbb{H-T}}\!\!\!\omega ^{(I,\xi )}.
\end{eqnarray*}%
Now note that $\left\Vert \mathbf{f}_{\mathbb{H}}-\mathbf{f}_{\mathbb{T}%
}\right\Vert _{1}=||\mathbf{f}_{\mathbb{H}}||_{1}-||\mathbf{f}_{\mathbb{T}}%
\mathbf{||}_{1}$, moreover, \allowdisplaybreaks%
\begin{eqnarray*}
&&\!\!\!\!\!\!\!\!\!\!\!\!\!\!\!\!\!\!\left\Vert \frac{\mathbf{f}_{\mathbb{H}%
}}{||\mathbf{f}_{\mathbb{H}}||_{1}}-\frac{\mathbf{f}_{\mathbb{T}}}{||\mathbf{%
f}_{\mathbb{T}}\mathbf{||}_{1}}\right\Vert _{1} \\
\!\! &\leq &\!\!\!\int \left\vert \Delta (\mathbf{X}\!)\!\!\!\sum%
\limits_{(I,\xi )\in \mathbb{T}}\!\!\left( \!\frac{\omega ^{(I,\xi )}}{||%
\mathbf{f}_{\mathbb{H}}||_{1}}\!-\!\frac{\omega ^{(I,\xi )}}{||\mathbf{f}_{%
\mathbb{T}}\mathbf{||}_{1}}\!\right) \!\delta _{I}(\mathcal{L(}\mathbf{X}))\!%
\left[ p^{(\xi )\!}\right] ^{\mathbf{X}}\right\vert \!\delta \mathbf{X} \\
&&\!\!\!+\int \left\vert \Delta (\mathbf{X})\sum\limits_{(I,\xi )\in \mathbb{%
H-T}}\frac{\omega ^{(I,\xi )}}{||\mathbf{f}_{\mathbb{H}}||_{1}}\delta _{I}(%
\mathcal{L(}\mathbf{X}))\left[ p^{(\xi )}\right] ^{\mathbf{X}}\right\vert
\delta \mathbf{X} \\
\!\! &=&\!\!\!\sum\limits_{(I,\xi )\in \mathbb{T}}\left\vert \frac{\omega
^{(I,\xi )}}{||\mathbf{f}_{\mathbb{H}}||_{1}}-\frac{\omega ^{(I,\xi )}}{||%
\mathbf{f}_{\mathbb{T}}\mathbf{||}_{1}}\right\vert +\sum\limits_{(I,\xi )\in
\mathbb{H-T}}\frac{\omega ^{(I,\xi )}}{||\mathbf{f}_{\mathbb{H}}||_{1}} \\
\!\! &=&\!\!\!1-\frac{||\mathbf{f}_{\mathbb{T}}\mathbf{||}_{1}}{||\mathbf{f}%
_{\mathbb{H}}||_{1}}+\frac{||\mathbf{f}_{\mathbb{H}}||_{1}-||\mathbf{f}_{%
\mathbb{T}}\mathbf{||}_{1}}{||\mathbf{f}_{\mathbb{H}}||_{1}} \\
\!\! &=&\!\!\!2\frac{||\mathbf{f}_{\mathbb{H}}||_{1}-||\mathbf{f}_{\mathbb{T}%
}\mathbf{||}_{1}}{||\mathbf{f}_{\mathbb{H}}||_{1}}. \square
\end{eqnarray*}

It follows from the above result that the intuitive strategy of keeping $%
\delta $-GLMB components with high weights and discarding those with the
smallest weights minimizes the $L_{1}$-error in the truncated multi-target
density.

In the $\delta $-GLMB recursion, it is not tractable to exhaustively compute
all the components first and then discard those with small weights. The
trick is to perform the truncation without having to propagate all the
components.

\section{Delta-GLMB Update \label{sec:delta-GLMB-MTT}}

This section presents a tractable implementation of the $\delta $-GLMB
update by truncating the multi-target filtering density without computing
all the hypotheses and their weights, via the ranked assignment algorithm.
Subsection \ref{subsec:ROAProblem} summarizes the ranked assignment problem
in the context of truncating the $\delta $-GLMB filtering density.
Subsection \ref{subsec:ComputingUpdateParam} details the computation of the
updated $\delta $-GLMB parameters and subsection \ref%
{subsec:posteriortruncation} presents the $\delta $-GLMB update algorithm.

\subsection{Ranked Assignment Problem\label{subsec:ROAProblem}}

Note from the $\delta $-GLMB weight update (\ref{eq:PropBayes_strong1}) that
each hypothesis $(I,\xi )$ with weight $\omega ^{(I,\xi )}$ generates a new
set of hypotheses $(I,(\xi ,\theta ))$, $\theta \in \Theta (I)$, with
weights $\omega ^{(I,\xi ,\theta )}(Z)$ $\propto $ $\omega ^{(I,\xi )}[\eta
_{Z}^{(\xi ,\theta )}]^{I}$. For a given hypothesis $(I,\xi )$, if we can
generate the association maps $\theta \in \Theta (I)$ in decreasing order of
$[\eta _{Z}^{(\xi ,\theta )}]^{I}$, then the highest weighted components can
be selected without exhaustively computing all the new hypothesis and their
weights. This can be accomplished by solving the following ranked assignment
problem.

Enumerating $Z=\{z_{1},...,z_{\left\vert Z\right\vert }\}$, $I=\{\ell
_{1},...,\ell _{\left\vert I\right\vert }\}$, each association map $\theta
\in \Theta (I)$ can be represented by an $\left\vert I\right\vert \times
\left\vert Z\right\vert $ \emph{assignment matrix} $S$ consisting of $0$ or $%
1$ entries with every row and column summing to either 1 or 0. For $i\in
\{1,...,\left\vert I\right\vert \}$, $j\in \{1,...,\left\vert Z\right\vert
\} $, $S_{i,j}=1$ if and only if the $j$th measurement is assigned to track $%
\ell _{i}$, i.e. $\theta (\ell _{i})$ $=$ $j$. An all-zero row $i$ means
that track $\ell _{i}$ is misdetected while an all-zero column $j$ means
that measurement $z_{j}$ is a false alarm. Conversion from $S$ to $\theta $
is given by $\theta (\ell _{i})=\sum_{j=1}^{\left\vert Z\right\vert }j\delta
_{1}(S_{i,j})$.

The \emph{cost matrix} of an optimal assignment problem is the $\left\vert
I\right\vert \times \left\vert Z\right\vert $ matrix:
\begin{equation}
C_{Z}^{(I,\xi )}=\left[
\begin{array}{ccc}
C_{1,1} & \cdots & C_{1,\left\vert Z\right\vert } \\
\vdots & \ddots & \vdots \\
C_{\left\vert I\right\vert ,1} & \cdots & C_{\left\vert I\right\vert
,\left\vert Z\right\vert }%
\end{array}%
\right]  \label{eq:Ass_Cost_Matrix}
\end{equation}%
where for $i\in \{1,...,\left\vert I\right\vert \}$, $j\in
\{1,...,\left\vert Z\right\vert \}$%
\begin{equation}
C_{i,j}=-\ln \left( \frac{\left\langle p^{(\xi )}(\cdot ,\ell
_{i}),p_{D}(\cdot ,\ell _{i})g(z_{j}|\cdot ,\ell _{i})\right\rangle }{%
\left\langle p^{(\xi )}(\cdot ,\ell _{i}),1-p_{D}(\cdot ,\ell
_{i})\right\rangle \kappa (z_{j})}\right)  \label{eq:Ass_Cost_Matrix1}
\end{equation}%
is the cost of the assigning the $j$th measurement to track $\ell _{i}$
(Subsection \ref{subsec:ComputingUpdateParam} details the
numerical computation of $C_{i,j}$).

The cost of an assignment (matrix) $S$ is the combined costs of every
measurement to target assignments, which can be succinctly written as the
Frobenius inner product
\begin{equation*}
\text{tr}(S^{T}C_{Z}^{(I,\xi )})=\sum_{i=1}^{\left\vert I\right\vert
}\sum_{j=1}^{\left\vert Z\right\vert }C_{i,j}S_{i,j}.
\end{equation*}%
Substituting (\ref{eq:PropConj5}) into equation (\ref{eq:PropBayes_strong2}%
), it follows that the cost of $S$ (and the corresponding association map $%
\theta $) is related to the filtered hypothesis weight $\omega ^{(I,\xi
,\theta )}(Z)$ $\propto $ $\omega ^{(I,\xi )}[\eta _{Z}^{(\xi ,\theta
)}]^{I} $ by
\begin{equation*}
\left[ \eta _{Z}^{(\xi ,\theta )}\right] ^{I}\!=\exp \!\left( -\text{tr}%
(S^{T}C_{Z}^{(I,\xi )})\!\right) \!\prod\limits_{\ell \in I}\!\left\langle
p^{(\xi )}(\cdot ,\ell ),1-p_{D}(\cdot ,\ell )\!\right\rangle .
\end{equation*}

The optimal assignment problem seeks an assignment matrix $S^{\ast }$ (and
corresponding association map $\theta ^{\ast }$) that minimizes the cost tr$%
(S^{\ast T}C_{Z}^{(I,\xi )})$ \cite{Kuhn55}. The ranked assignment problem
seeks an enumeration of the least cost assignment matrices in non-decreasing
order \cite{Murty}. Consequently, solving the ranked optimal assignment
problem with cost matrix $C_{Z}^{(I,\xi )}$ generates, starting from $\theta
^{\ast }$, an enumeration of association maps $\theta $ in order of
non-increasing $[\eta _{Z}^{(\xi ,\theta )}]^{I}$ (and weights $\omega
^{(I,\xi ,\theta )}(Z)$ $\propto $ $\omega ^{(I,\xi )}[\eta _{Z}^{(\xi
,\theta )}]^{I}$).

Remark: The standard ranked assignment formulation involves square cost and
assignment matrices with rows and columns of the assignment matrix summing
to 1. Ranked assignment problems with non-square matrices can be reformulated with square matrices by introducing dummy variables.

The optimal assignment problem, is a well-known combinatorial problem,
introduced by Kuhn \cite{Kuhn55}, who also proposed the Hungarian algorithm
to solve it in polynomial time. Munkres further observed that it is strongly
polynomial \cite{Munkres57}. The ranked assignment problem is a
generalization to enumerate the $T$ least cost assignments, which was first
solved by Murty \cite{Murty}. Murty's algorithm needs an effective bipartite
assignment algorithm such as Munkres \cite{Munkres57} or Jonker-Volgenant
\cite{Jonker87}. In the context of multi-target tracking, ranked assignment
algorithms with $O(T\left\vert Z\right\vert ^{4})$ complexity have been
proposed for MHT in \cite{DanchickNewnam93}, \cite{CoxMiller95}, \cite%
{CoxHingorani96}. More efficient algorithms with $O(T\left\vert Z\right\vert
^{3})$ complexity have been proposed in \cite{Milleretal97}, \cite%
{Pascoaletal03}, \cite{Pedersenetal08}, with the latter showing better
efficiency for large $\left\vert Z\right\vert $. For further details on
ranked assignment solutions in MHT, we refer the reader to \cite{Blackman,
Pattipati}.

\subsection{Computing update parameters\label{subsec:ComputingUpdateParam}}

We now detail the computation of the cost matrix $C_{Z}^{(I,\xi )}$ in (\ref%
{eq:Ass_Cost_Matrix}) for the ranked assignment problem and the parameters $%
\eta _{Z}^{(\xi ,\theta )}(\ell )$, $p^{\!(\xi ,\theta )\!}(\cdot ,\ell |Z)$
of the updated $\delta $-GLMB components.

\subsubsection{Gaussian mixture}

For a linear Gaussian multi-target model, $p_{D}(x,\ell )=p_{D}$, $%
g(z|x,\ell )=\mathcal{N}(z;Hx,R)$, where $\mathcal{N}(\cdot ;m,P)$ denotes a
Gaussian density with mean $m$ and covariance $P$, $H$ is the observation
matrix, and $R$ is the observation noise covariance. The Gaussian mixture
representation provides the most general setting for linear Gaussian models.
Suppose that each single target density $p^{(\xi )}(\cdot ,\ell )$ is a
Gaussian mixture of the form%
\begin{equation}
\sum_{i=1}^{J^{(\xi )}(\ell )}w_{i}^{(\xi )}(\ell )\mathcal{N}(x;m_{i}^{(\xi
)}(\ell ),P_{i}^{(\xi )}(\ell )).  \label{eq:GM_single_pdf}
\end{equation}%
Then%
\begin{equation}
C_{i,j}=-\ln \left( \frac{p_{D}\sum_{k=1}^{J^{(\xi )}(\ell
_{i})}w_{k}^{\!(\!\xi )}\!(\ell _{i})q_{k}^{\!(\!\xi )}\!(z_{j};\ell _{i})}{%
(1-p_{D})\kappa (z_{j})}\right)   \label{eq:GM_Ass_Cost_Matrix1}
\end{equation}%
Moreover, for the updated association history $(\xi ,\theta )$,
\begin{eqnarray}
\!\!\!\!\!\!\eta _{Z}^{(\xi ,\theta )}(\ell )\!\!\!
&=&\!\!\!\!\sum\limits_{i=1}^{J^{(\xi )}(\ell )}w_{Z,i}^{\!(\xi ,\theta
)}\!(\ell )  \label{eq:GM_eta_Z} \\
\!\!\!\!\!\!p^{\!(\xi ,\theta )\!}(x,\ell |Z)\!\!\!
&=&\!\!\!\!\sum\limits_{i=1}^{J^{(\xi )}\!(\ell )}\!\!\frac{w_{Z,i}^{\!(\xi
,\theta )}\!(\ell )}{\eta _{Z}^{(\xi ,\theta )}(\ell )}\mathcal{N}%
\!(x;m_{Z,i}^{\!(\xi ,\theta )}\!(\ell ),P_{i}^{\!(\xi ,\theta )}\!(\ell ))
\label{eq:GM_single_update}
\end{eqnarray}%
where{%
\begin{align*}
w_{Z,i}^{\!(\!\xi ,\theta )}\!(\ell )& =w_{i}^{\!(\!\xi )}\!(\ell )\left\{
\!\!\!%
\begin{array}{ll}
\frac{p_{D}q_{i}^{\!(\!\xi )}\!(z_{\theta (\ell )};\ell )}{\kappa (z_{\theta
(\ell )})} & \text{if }\theta (\ell )>0 \\
(1-p_{D}) & \text{if }\theta (\ell )=0%
\end{array}%
\right.  \\
q_{i}^{\!(\!\xi )}\!(z;\ell )& =\mathcal{N}\!(z;\!Hm_{i}^{\!(\xi )}\!(\ell
),HP_{i}^{\!(\xi )}\!(\ell )H^{T}\!+\!R) \\
m_{Z,i}^{\!(\!\xi ,\theta )}\!(\ell )& =\left\{ \!\!\!%
\begin{array}{ll}
m_{i}^{\!(\xi )}\!(\ell )\!+\!K_{i}^{(\xi ,\theta )}\!(\ell )(z_{\theta
(\ell )}\!-\!Hm_{i}^{\!(\xi )}\!(\ell )) & \!\!\!\!\text{if }\theta (\ell )>0
\\
m_{i}^{\!(\xi )}\!(\ell ) & \!\!\!\!\text{if }\theta (\ell )=0%
\end{array}%
\right.  \\
P_{i}^{(\!\xi ,\theta )}\!(\ell )& =[I-K_{i}^{(\xi ,\theta )}\!(\ell
)H]P_{i}^{(\xi )}\!(\ell ), \\
K_{i}^{(\!\xi ,\theta )}\!(\ell )& =\left\{ \!\!\!%
\begin{array}{ll}
P_{i}^{(\xi )}\!(\ell )H^{T}\!\!\left[ \!H\!P_{i}^{(\xi )}\!(\ell
)H^{T}\!+\!R\right] ^{-1} & \!\!\text{if }\theta (\ell )>0 \\
0 & \!\!\text{if }\theta (\ell )=0%
\end{array}%
\right. .
\end{align*}%
}When the measurement model parameters depend on the label $\ell $, we
simply substitute $p_{D}=p_{D}(\ell ),H=H(\ell ),R=R(\ell )$ into the above
equations.

\subsubsection{Sequential Monte Carlo}

For a sequential Monte Carlo approximation, suppose that each of the single
target density $p^{(\xi )}(\cdot ,\ell )$ is represented as a set of
weighted samples $\{(w_{n}^{(\xi )}(\ell ),x_{n}^{(\xi )}(\ell
))\}_{n=1}^{J^{(\xi )}(\ell )}$. Then%
\begin{equation}
C_{i,j}=-\ln \left( \!\frac{\sum\limits_{n=1}^{J^{(\xi )}\!(\ell
_{i})}\!w_{n}^{(\xi )}\!(\ell _{i})p_{D}(x_{n}^{(\xi )}\!(\ell _{i\!}),\ell
_{i\!})g(z_{j}|x_{n}^{(\xi )}\!(\ell _{i\!}),\ell _{i\!})}{%
\sum\limits_{n=1}^{J^{(\xi )}\!(\ell _{i})}\!w_{n}^{(\xi )}\!(\ell
_{i\!})(1-p_{D}(x_{n}^{(\xi )}\!(\ell _{i\!}),\ell _{i\!}))\kappa (z_{j})}%
\!\right) .  \label{eq:Particle_Ass_Cost_Matrix1}
\end{equation}%
Moreover, for a given updated association history $(\xi ,\theta )$,%
\begin{equation}
\eta _{Z}^{(\xi ,\theta )}(\ell )=\sum\limits_{n=1}^{J^{(\xi )}(\ell
)}w_{n}^{(\xi )}(\ell )\psi _{Z}(x_{n}^{(\xi )}\!(\ell ),\ell ;\theta ),
\label{eq:Particle_eta_Z}
\end{equation}%
and $p^{\!(\xi ,\theta )\!}(\cdot ,\ell |Z)$ is represented by the following
set of weighted samples%
\begin{equation}
\left\{ \left( \frac{\psi _{Z}(x_{n}^{(\xi )}\!(\ell ),\ell ;\theta
)w_{n}^{(\xi )}(\ell )}{\eta _{Z}^{(\xi ,\theta )}(\ell )},x_{n}^{(\xi
)}(\ell )\right) \right\} _{n=1}^{J^{(\xi )}(\ell )}.
\label{eq:Particle_single_update}
\end{equation}

\subsection{Filtering Density Truncation\label{subsec:posteriortruncation}}

Given the $\delta $-GLMB prediction density $\mathbf{\pi }$ with enumerated
parameter set $\{(I^{(h)},\xi ^{(h)},\omega ^{(h)},p^{(h)})\}_{h=1}^{H}$,
the $\delta $-GLMB filtering density (\ref{eq:PropBayes_strong0}) can be
written as
\begin{equation}
\mathbf{\pi \!}(\mathbf{X}|Z)=\sum_{h=1}^{H}\mathbf{\pi }^{(h)}\mathbf{\!}(%
\mathbf{X}|Z)  \label{eq:posterior_param}
\end{equation}%
where%
\begin{eqnarray*}
\mathbf{\pi }^{(h)}\mathbf{\!}(\mathbf{X}|Z) &=&\mathbf{\!}\Delta (\mathbf{X}%
_{\mathbf{\!}})\mathbf{\!\!\!}\sum\limits_{j=1}^{\left\vert \Theta (I^{(h)%
\mathbf{\!}})\mathbf{\!}\right\vert }\mathbf{\!\!}\omega ^{(h,j)}\delta _{%
\mathbf{\!}I^{(h\mathbf{\!})}}\mathbf{\!}(\mathcal{L(}\mathbf{X}))\mathbf{\!}%
\left[ p^{(h,j)}\mathbf{\!}\right] ^{\mathbf{X}}, \\
\omega ^{(h,j)} &\triangleq &\omega ^{(I^{(h\mathbf{\!})},\xi ^{(h\mathbf{\!}%
)},\theta ^{(h,j)})}(Z), \\
p^{(h,j)} &\triangleq &p^{(\xi ^{(h\mathbf{\!})},\theta ^{(h,j)})}(\cdot |Z).
\end{eqnarray*}%
Each $\delta $-GLMB prediction component (indexed by) $h$ generates
$\left\vert \Theta (I^{(h)})\right\vert $ components for the $\delta $-GLMB
filtering density.

\begin{figure}[tbh]
\begin{center}
\resizebox{80mm}{!}{\includegraphics[clip=true]{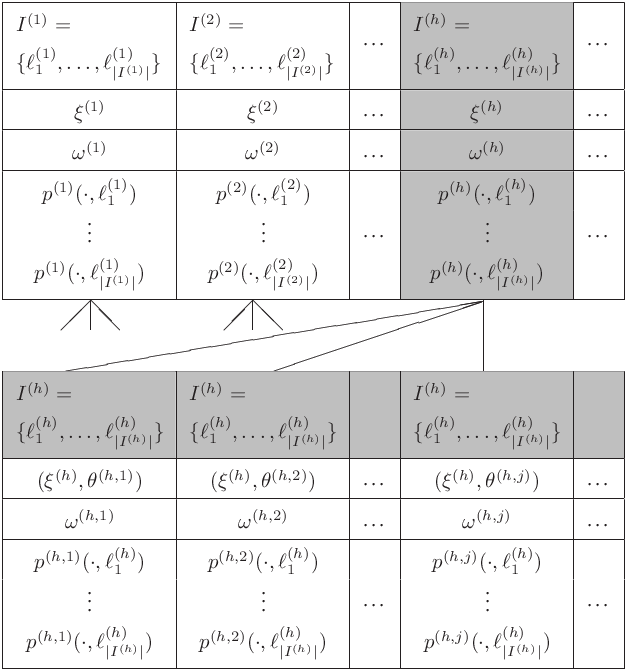}}
\end{center}
\caption{$\protect\delta $-GLMB update. Component $h$ of the prior generates
a (large) set of posterior components. The ranked assignment algorithm
determines the $T^{(h)}$ components with highest weights $\protect\omega %
^{(h,1)}\geq \protect\omega ^{(h,2)}\geq ...\geq \protect\omega %
^{(h,T^{(h)})}$.}
\label{fig:updatecrop}
\end{figure}

A simple and highly parallelizable strategy for truncating the filtered $%
\delta $-GLMB (\ref{eq:posterior_param}) is to truncate $\mathbf{\pi }_{%
\mathbf{\!}}^{(h)}(\mathbf{\cdot }|Z)$. For each $h=1,...,H$, solving the
ranked optimal assignment problem with cost matrix $C_{Z}^{(I^{(h)},\xi
^{(h)})}$ discussed in subsection \ref{subsec:ROAProblem} yields $\theta
^{(h,j)}$, $j=1,...,T^{(h)}$, the $T^{(h)}$ hypotheses with highest weights
in non-increasing order, as illustrated in Figure \ref{fig:updatecrop}.
Consequently, the truncated version of $\mathbf{\pi }_{\mathbf{\!}}^{(h)}(%
\mathbf{\cdot }|Z)$ is
\begin{equation*}
\mathbf{\hat{\pi}}^{(h)}(\mathbf{X}|Z)=\Delta (\mathbf{X})\sum%
\limits_{j=1}^{T^{(h)}}\omega ^{(h,j)}\delta _{I^{(h\mathbf{\!})}}(\mathcal{%
L(}\mathbf{X}))\left[ p^{(h,j)}\right] ^{\mathbf{X}}.
\end{equation*}%
It follows from Proposition \ref{Prop_L1_error} that the truncated density $%
\mathbf{\hat{\pi}}(\mathbf{\cdot }|Z)=\sum_{h=1}^{H}\mathbf{\hat{\pi}}^{(h)}(%
\mathbf{\cdot }|Z)$ minimizes the $L_{1}$-distance from the filtered $\delta
$-GLMB over all truncations with $T^{(h)}$ components for each $h=1,...,H$.
The truncated density, with a total of $T=\sum_{h=1}^{H}T^{(h)}$ components,
is normalized (by the sum of the weights) to give the truncated filtered $%
\delta $-GLMB. Table 1 summarizes the update operation via pseudo code. Note
that both the outer and inner for-loops can be parallelized.

\begin{figure}[tbp]
\hrule
\par
\medskip
\par
\textbf{Table 1. Update}
\par
\begin{itemize}
\item {\footnotesize \textsf{input: }} $\{(I^{(h)},\xi ^{(h)},\omega
^{(h)},p^{(h)},T^{(h)})\}_{h=1}^{H}$, $Z$%
\par
\item \textsf{{\footnotesize {output: }}}$\{(I^{(h,j)},\xi ^{(h,j)},\omega
^{(h,j)},p^{(h,j)})\}_{(h,j)=(1,1)}^{(H,T^{(h)})}$%
\end{itemize}
\par
\hrule
\par
\medskip
\par
\textsf{{\footnotesize {for}}}\textsf{\ }$h=1:H$%
\par
\quad $C_{Z}^{(h)}:=C_{Z}^{(I^{(h)},\xi ^{(h)})}$ \ \textsf{{\footnotesize {%
according to}}} (\ref{eq:Ass_Cost_Matrix}), (\ref{eq:GM_Ass_Cost_Matrix1})/(%
\ref{eq:Particle_Ass_Cost_Matrix1})
\par
\quad $\{\theta ^{(h,j)}\}_{_{j=1}}^{T^{(h)}}:=\mathsf{ranked\_assignment}%
(Z,I^{(h)},C_{Z}^{(h)},T^{(h)})$%
\par
\quad \textsf{{\footnotesize {for }}}$j=1:T^{(h)}$%
\par
\quad \quad $\eta _{Z}^{(h,j)}:=\eta _{Z}^{(\xi ^{(h)},\theta ^{(h,j)})}$ \
\textsf{{\footnotesize {according to}}} (\ref{eq:GM_eta_Z})/(\ref%
{eq:Particle_eta_Z})
\par
\quad \quad $p^{(h,j)}:=p^{(\xi ^{(h)},\theta ^{(h,j)})}(\cdot |Z)$ \
\textsf{{\footnotesize {according to}}} (\ref{eq:GM_single_update})/(\ref%
{eq:Particle_single_update})
\par
\quad \quad $\omega ^{(h,j)}:=\omega ^{(h)}\left[ \eta _{Z}^{(h,j)}\right]
^{I^{(h)}}$%
\par
\quad \quad $I^{(h,j)}:=I^{(h)}$%
\par
\quad \quad $\xi ^{(h,j)}:=(\xi ^{(h)},$$\theta ^{(h,j)})$%
\par
\quad \textsf{{\footnotesize {end}}}
\par
\textsf{{\footnotesize {end}}}
\par
\textsf{{\footnotesize {normalize weights }}}$\{\omega
^{(h,j)}\}_{(h,j)=(1,1)}^{(H,T^{(h)})}$%
\par
\medskip
\par
\hrule
\end{figure}

Specific values for the number of requested components $T^{(h)}$ are
generally user specified and application dependent. A generic strategy is to
choose $T^{(h)}=\left\lceil \omega ^{(h)}J_{\max }\right\rceil $ where $%
J_{\max }$ is the desired overall number of hypotheses. The alternative
strategy of keeping the $T=J_{\max}$ strongest components of $\mathbf{\pi \!}%
(\mathbf{\cdot }|Z_{\mathbf{\!}})$ would yield a smaller $L_{1}$-error.
However, in addition to an $H$-fold increase in the dimension of the
resulting ranked assignment problem, parallelizability is lost.

\section{Delta-GLMB Prediction\label{sec:Prediction}}

This subsection presents an implementation of the $\delta $-GLMB prediction
using the $K$-shortest path algorithm to truncate the predicted $\delta $%
-GLMB without computing all the prediction hypotheses and their weights.

The prediction density given in Proposition \ref{Prop_CK_strong} has a
compact form but is difficult to implement due to the sum over all supersets
of $L$ in (\ref{eq:PropCKstrongws}). We use an equivalent form for the
prediction, eq. (58) in \cite{VoConj13}:%
\begin{eqnarray}
\mathbf{\pi }_{\mathbf{\!}+\mathbf{\!}}(\mathbf{X}_{\mathbf{\!}+\mathbf{\!}})%
\mathbf{\!} &=&\mathbf{\!}\Delta (\mathbf{X}_{\mathbf{\!}+\mathbf{\!}})%
\mathbf{\!\!}\sum_{(I,\xi )\in \mathcal{F}(\mathbb{L})\times \Xi }\mathbf{%
\!\!\!\!}\omega ^{(I,\xi )}\mathbf{\!\!\!}\sum_{J\in \mathcal{F}(I)}\mathbf{%
\!\!}[\eta _{S}^{(\xi )}\mathbf{]}^{J}[1-\eta _{S}^{(\xi )}\mathbf{\!}]^{I-J}%
\mathbf{\!}  \notag \\
&&\times \sum_{L\in \mathcal{F}(\mathbb{B})}\mathbf{\!}w_{B}(L)\delta
_{J\cup L\mathbf{\!}}(\mathcal{L(}\mathbf{X}_{+}))\left[ p_{+}^{(\xi )}%
\mathbf{\!}\right] ^{\mathbf{X}_{+}},  \label{eq:alternative_pred}
\end{eqnarray}%
Note that analogous to the update, each current hypothesis $(I,\xi )$ with
weight $\omega ^{(I,\xi )}$ generates a set of prediction hypotheses $(J\cup
L,\xi )$, $J\subseteq I$, $L\subseteq \mathbb{B}$, with weights $\omega
_{S}^{(I,\xi )}(J)w_{B}(L)$, where%
\begin{equation}
\omega _{S}^{(I,\xi )}(J)=\omega ^{(I,\xi )}[\eta _{S}^{(\xi )}\mathbf{]}%
^{J}[1-\eta _{S}^{(\xi )}]^{I-J}  \label{eq:omegha_S}
\end{equation}

Intuitively, each predicted label set $J\cup L$ consists of a surviving
label set $J$ with weight $\omega _{S}^{(I,\xi )}(J)$ and a birth label set $%
L$ with weight $w_{B}(L)$. The weight $\omega _{S}^{(I,\xi )}(J)$ can be
interpreted as the probability that the current label set is $I$, and the
labels in $J$ survives to the next time while the remaining labels $I-J$
die. The birth label set $L$ and the surviving label set $J$ are mutually
exclusive since the space of new labels $\mathbb{B}$ cannot contain any
existing labels. Since the weight of $J\cup L$ is the product $\omega
_{S}^{(I,\xi )}(J)w_{B}(L)$, \ we can truncate the double sum over $J$ and $%
L $ by separately truncating the sum over $J$ and the sum over $L$.

Subsection \ref{subsec:KSPProblem} discusses the $K$-shortest paths problem
in the context of truncating the $\delta $-GLMB prediction. Subsection \ref%
{subsec:ComputingPredParam} details the computation of the prediction $%
\delta $-GLMB parameters and subsection \ref{subsec:predtruncation} presents
the $\delta $-GLMB prediction algorithm.

\subsection{K-Shortest Paths Problem\label{subsec:KSPProblem}}

\begin{figure}[tbh]
\begin{center}
\resizebox{70mm}{!}{\includegraphics[clip=true]{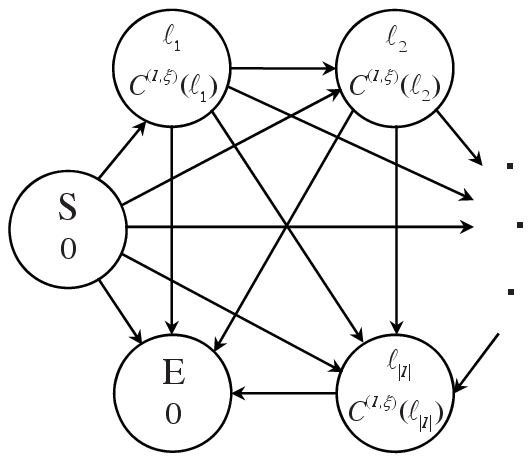}}
\end{center}
\caption{A directed graph with nodes $\ell _{1}$,...,$\ell _{\left\vert
I\right\vert }\in I$, and corresponding costs $C^{(I,\protect\xi )}(\ell
_{1})$,...,$C^{(I,\protect\xi )}(\ell _{\left\vert I\right\vert })$. S and E
are the start and end nodes respectively.}
\label{fig:kshortest}
\end{figure}

Consider a given hypothesis $(I,\xi )$, and note that the weight of a
surviving label set $J\subseteq I$ can be rewritten as%
\begin{equation*}
\omega _{S}^{(I,\xi )}(J)=\omega ^{(I,\xi )}[1-\eta _{S}^{(\xi )}]^{I}\left[
\frac{\eta _{S}^{(\xi )}}{1-\eta _{S}^{(\xi )}}\right] ^{J}
\end{equation*}%
If we can generate the surviving label sets $J\subseteq I$ in non-increasing
order of $[\eta _{S}^{(\xi )}/(1-\eta _{S}^{(\xi )})\mathbf{]}^{J}$, then
the highest weighted survival sets for hypothesis $(I,\xi )$ can be selected
without exhaustively computing all the survival hypotheses weights. This can
be accomplished by solving the $K$-shortest path problem in the directed
graph of Figure \ref{fig:kshortest}.

Define a cost vector $C^{(I,\xi )}=\left[ C^{(I,\xi )}(\ell
_{1}),...,C^{(I,\xi )}(\ell _{\left\vert I\right\vert })\right] $, where
\begin{equation}
C^{(I,\xi )}(\ell _{j})=-\ln \left[ \frac{\eta _{S}^{(\xi )}(\ell _{j})}{%
1-\eta _{S}^{(\xi )}(\ell _{j})}\right]  \label{eq:costKshortest}
\end{equation}%
is the cost of node $\ell _{j}\in I$ (the numerical computation of $%
C^{(I,\xi )}(\ell _{j})$ is detailed in Subsection \ref%
{subsec:ComputingPredParam}). The nodes are ordered in non-decreasing costs
and the distance from node $\ell _{j}$ to $\ell _{j^{\prime }}$\ is defined
as
\begin{equation*}
d(\ell _{j},\ell _{j^{\prime }})=\left\{
\begin{array}{ll}
C^{(I,\xi )}(\ell _{j^{\prime }}), & \text{if }j^{\prime }>j \\
\infty , & \text{otherwise}%
\end{array}%
\right.
\end{equation*}%
Hence, a path from S to E which traverses the set of nodes $J\subseteq I$
accumulates a total distance of%
\begin{eqnarray*}
\sum_{\ell \in J}C^{(I,\xi )}(\ell ) &=&-\sum_{\ell \in J}\ln \left( \eta
_{S}^{(\xi )}(\ell )/(1-\eta _{S}^{(\xi )}(\ell ))\right) \\
&=&-\ln \left( [\eta _{S}^{(\xi )}(\ell )/(1-\eta _{S}^{(\xi )}(\ell
)]^{J}\right) .
\end{eqnarray*}

The shortest path from S to E traverses the set of nodes $J^{\ast }\subseteq
I$ with the shortest distance $\sum_{\ell \in J^{\ast }}C^{(I,\xi )}(\ell )$
and hence largest $[\eta _{S}^{(\xi )}/(1-\eta _{S}^{(\xi )})]^{J^{\ast }}$.
The $K$-shortest paths problem seeks $K$ subsets of $I$ with the shortest
distances in non-decreasing order. Consequently, solving the $K$-shortest
path problem generates, starting from $J^{\ast }$, an enumeration of subsets
$J$ of $I$ in order of non-increasing $[\eta _{S}^{(\xi )}/(1-\eta
_{S}^{(\xi )})]^{J}$.

For the target births we use a labeled multi-Bernoulli birth model where
\begin{eqnarray*}
w_{B}(L) &=&\prod\limits_{\ell \in \mathbb{B}}\left( 1-r_{B}^{(\ell
)}\right) \prod\limits_{\ell \in L}\frac{1_{\mathbb{B}}(\ell )r_{B}^{(\ell )}%
}{1-r_{B}^{(\ell )}}, \\
p_{B}(x,\ell ) &=&p_{B}^{(\ell )}(x).
\end{eqnarray*}%
Thus, solving the $K$-shortest paths problem with cost vector $C_{B}=\left[
C_{B}(\ell _{1}),...,C_{B}(\ell _{\left\vert \mathbb{B}\right\vert })\right]
$, where
\begin{equation}
C_{B}(\ell _{j})=-\ln \left[ r_{B}^{(\ell _{j})}/(1-r_{B}^{(\ell _{j})})%
\right] \   \label{eq:costvectorbirth}
\end{equation}%
is the cost of node $\ell _{j}$, yields the subsets of $\mathbb{B}$ with the
best birth weights.

Remark: It is possible to obtain the overall $K$ best components by
extending the directed graphs to include birth nodes with appropriate costs.
However, our experience indicated that since the birth weights $w_{B}(L)$
are quite small compared to surviving weights $\omega _{S}^{(I,\xi )}(J)$,
many components with births will be discarded and new births may not
detected by the filter. To avoid dropping new tracks, a very large $K$ is
required to retain hypothesis with births. On the other hand the proposed
separate truncation strategy ensures that there are hypotheses with births
to accommodate new tracks, and is highly parallelizable.

The $K$-shortest paths algorithm is a well-known solution to the
combinatorial problem of finding the $K$ paths with minimum total cost from
a given source to a given destination in a weighted network \cite%
{Eppstein98finding}. This problem can be solved with complexity $O\left(
\left\vert I\right\vert \log (\left\vert I\right\vert )+K\right) $. In our
case the nodes have negative values, hence the Bellman-Ford algorithm \cite%
{Bellman52routing, FordFulkersonbk62} was employed. This problem can also be
solved by ranked assignment algorithms, however $K$-shortest paths algorithm
is much more efficient. Note that the ranking of the association maps cannot
be formulated as the $K$-shortest path problem, due to the constraint that
each target can only generate at most one measurement.

\subsection{Computing prediction parameters\label{subsec:ComputingPredParam}}

This subsection details the computation of the parameters $\eta _{S}^{(\xi
)}(\ell )$ and $p_{+}^{(\xi )}(\cdot ,\ell )$ of the prediction $\delta $%
-GLMB components.

\subsubsection{Gaussian mixture}

For a linear Gaussian multi-target model, $p_{S}(x,\ell )=p_{S}$, $%
f(x_{+}|x,\ell )=\mathcal{N}(x_{+};Fx,Q)$, where $F$ is the state transition
matrix, $Q$ is the process noise covariance and the birth density parameter $%
p_{B}^{(\ell )}(x)$ is a Gaussian mixture. If the single target density $%
p^{(\xi )}(\cdot ,\ell )$ is a Gaussian mixture of the form (\ref%
{eq:GM_single_pdf}). Then,%
\begin{eqnarray}
\eta _{S}^{(\xi )\!}(\ell )\!\!\!\! &=&\!\!\!\!p_{S},  \label{eq:GM_eta_S} \\
p_{+}^{(\xi )\!}(x,\ell )\!\!\!\! &=&\!\!\!\!1_{\mathbb{L}}\!(\ell
)\!\!\!\sum_{i=1}^{J^{(\xi )}\!(\ell \!)}\!\!w_{i}^{\!(\xi )}\!(\ell )%
\mathcal{N}\!(x;m_{\!S,i}^{\!(\xi )}\!(\ell )\!,P_{\!S,i}^{(\xi )}\!(\ell
)\!)\!+\!1_{\mathbb{B}}\!(\ell )p_{B}^{(\ell )}\!(x)  \notag \\
&&  \label{eq:GM_single_predict}
\end{eqnarray}%
where
\begin{eqnarray*}
m_{S,i}^{(\xi )}(\ell ) &=&Fm_{i}^{(\xi )}(\ell ), \\
P_{S,i}^{(\xi )}(\ell ) &=&Q+FP_{i}^{(\xi )}(\ell )F^{T}.
\end{eqnarray*}%
When the motion model parameters depend on the label $\ell $, we simply
substitute $p_{S}=p_{S}(\ell )$, $F=F(\ell )$, $Q=Q(\ell )$ into the above
equations. With $\eta _{S}^{(\xi )}(\ell )$ evaluated, the node cost $%
C^{(I,\xi )}(\ell )$ for the $K$-shortest paths problem can then be computed
by (\ref{eq:costKshortest}).

\subsubsection{Sequential Monte Carlo}

For a sequential Monte Carlo approximation, suppose that each single target
density $p^{(\xi )}(\cdot ,\ell )$ is represented as a set of weighted
sample $\{(w_{i}^{(\xi )}(\ell ),x_{i}^{(\xi )}(\ell ))\}_{i=1}^{J^{(\xi
)}(\ell )}$ and that the birth density $p_{B}^{(\ell )}(\cdot )$ is
represented by $\{(w_{B,i}^{(\xi )}(\ell ),x_{B,i}^{(\xi )}(\ell
))\}_{i=1}^{B^{(\xi )}(\ell )}$. Then,%
\begin{equation}
\eta _{S}^{(\xi )}(\ell )=\sum\limits_{i=1}^{J^{(\xi )}(\ell )}w_{i}^{(\xi
)}(\ell )p_{S}(x_{i}^{(\xi )}(\ell ),\ell )  \label{eq:Particle_eta_S}
\end{equation}%
and $p_{+}^{(\xi )}(x,\ell )$ is represented by
\begin{equation}
\left\{ \!(1_{\mathbb{L}\!}(\ell )\tilde{w}_{\!S,i}^{\!(\xi )}\!(\ell
),x_{\!S,i}^{\!(\xi )}\!(\ell ))\!\right\} _{i=1}^{\!J^{(\xi )\!}(\ell
)}\!\cup \!\left\{ \!(1_{\mathbb{B}\!}(\ell )w_{\!B,i}^{\!(\xi )}\!(\ell
),x_{\!B,i}^{\!(\xi )}\!(\ell ))\!\right\} _{i=1}^{\!B^{(\xi )\!}(\ell )},
\label{eq:Particle_single_predict}
\end{equation}%
where%
\begin{eqnarray*}
x_{S,i}^{(\xi )}(\ell )\!\!\! &\sim &\!\!\!q^{(\xi )}(\cdot |x_{i}^{(\xi
)}(\ell ),\ell ,Z),~i=1,...,J^{(\xi )}(\ell ), \\
w_{S,i}^{(\xi )}(\ell )\!\!\! &=&\!\!\!\frac{w_{i}^{(\xi )}(\ell
)f(x_{S,i}^{(\xi )}(\ell )|x_{i}^{(\xi )}(\ell ))p_{S}(x_{i}^{(\xi )}(\ell
),\ell )}{q^{(\xi )}(x_{S,i}^{(\xi )}(\ell )|x_{i}^{(\xi )}(\ell ),\ell ,Z)},
\\
\tilde{w}_{S,i}^{(\xi )}(\ell )\!\!\! &=&\!\!\!\frac{w_{S,i}^{(\xi )}(\ell
)\!}{\sum_{i=1}^{J^{(\xi )}(\ell )}w_{S,i}^{(\xi )}(\ell )},
\end{eqnarray*}%
and $q^{(\xi )}(\cdot |x_{i}^{(\xi )}(\ell ),\ell ,Z)$ is a proposal density.

\subsection{Prediction Density Truncation\label{subsec:predtruncation}}

Given the current $\delta $-GLMB filtering density $\mathbf{\pi }(\mathbf{%
\cdot }|Z)$ with enumerated parameter set $\{(I^{(h)},\xi ^{(h)},\omega
^{(h)},p^{(h)})\}_{h=1}^{H}$, then the $\delta $-GLMB prediction (\ref%
{eq:alternative_pred}) becomes%
\begin{equation*}
\mathbf{\pi }_{+}(\mathbf{X}_{+})=\sum_{h=1}^{H}\mathbf{\pi }_{+}^{(h)}(%
\mathbf{X}_{+}),
\end{equation*}%
where
\begin{eqnarray*}
\mathbf{\pi }_{\mathbf{\!}+\mathbf{\!}}^{(h)}\mathbf{\!}(\mathbf{X}_{\mathbf{%
\!}+\mathbf{\!}})\mathbf{\!\!} &=& \\
&&\mathbf{\!\!\!\!\!\!\!\!\!\!\!\!\!\!\!\!\!\!\!\!\!\!\!\!\!\!\!\!\!\!\!\!\!}%
\Delta \mathbf{\!}(\mathbf{X}_{\mathbf{\!}+})\mathbf{\!\!}%
\sum\limits_{J\subseteq I^{(\mathbf{\!}h\mathbf{\!})\mathbf{\!}%
}}\sum_{L\subseteq \mathbb{B}}\mathbf{\!\!}\omega _{S}^{(I^{(h)}\mathbf{\!}%
,\xi ^{(h)\mathbf{\!}})}\mathbf{\!}(\mathbf{\!}J)w_{B\mathbf{\!}}(L\mathbf{\!%
})\delta \mathbf{\!}_{J\cup L\mathbf{\!}}(\mathbf{\!}\mathcal{L(}\mathbf{X}%
_{+\mathbf{\!}}))\mathbf{\!\!}\left[ p_{+}^{(\xi ^{(h)\mathbf{\!}})}\mathbf{%
\!}\right] ^{\mathbf{\!X}_{\mathbf{\!}+}}.
\end{eqnarray*}%
The $\delta $-GLMB filtered component (indexed by) $h$ generates $%
2^{\left\vert I^{(h)\mathbf{\!}}\right\vert +\left\vert \mathbb{B}%
\right\vert }$components\ for the $\delta $-GLMB prediction.

\begin{figure}[tbh]
\begin{center}
\resizebox{85mm}{!}{\includegraphics[clip=true]{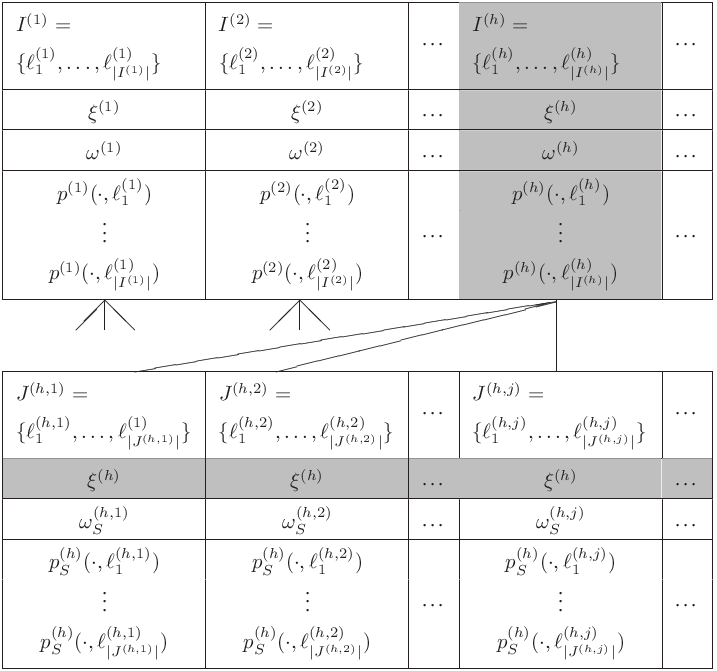}}
\end{center}
\caption{Prediction of survival components. Component $h$ of the prior,
generates all subsets of $I^{(h)}$ , i.e. $J^{(h,j)},j=1$,..., $%
2^{|I^{(h)}|} $ with weights $\protect\omega _{S}^{(h,j)}\triangleq \protect%
\omega _{S}^{(I^{(h)},\protect\xi ^{(h)})}(J^{(h,j)})$. The $K$-shortest
paths algorithm determines the $K^{(h)}$ subsets with largest weights $%
\protect\omega _{S}^{(h,1)}\geq \protect\omega _{S}^{(h,2)}\geq ...\geq
\protect\omega _{S}^{(h,K^{(h)})}$.}
\label{fig:predsurvivecrop}
\end{figure}

A simple and highly parallelizable strategy for truncating the prediction $%
\delta $-GLMB $\mathbf{\pi }_{+}$ is to truncate each $\mathbf{\pi }_{%
\mathbf{\!}+\mathbf{\!}}^{(h)}$ as follows. For each $h=1,...,H$, we solve
the $K$-shortest paths problem with cost vector $C^{(I^{(h)},\xi ^{(h)})}$
to obtain $J^{(h,j)},j=1,...,K^{(h)},$ the $K^{(h)}$ subsets of $I^{\mathbf{%
\!}(h\mathbf{\!})}$ with highest survival weights as depicted in Figure \ref%
{fig:predsurvivecrop}. We also solve the $K$-shortest paths problem with
cost vector $C_{B}$ to obtain $L^{(b)},b=1,...,K_{B}$, the $K_{B}$ birth
subsets with highest birth weights. Consequently for each $h$, the truncated
version of $\mathbf{\pi }_{\mathbf{\!}+\mathbf{\!}}^{(h)}$ is
\begin{equation*}
\mathbf{\hat{\pi}}_{\mathbf{\!}+\mathbf{\!}}^{(h)}\mathbf{\!}(\mathbf{X}_{%
\mathbf{\!}+\mathbf{\!}})\mathbf{\!}=\mathbf{\!}\Delta \mathbf{\!}(\mathbf{X}%
_{\mathbf{\!}+})\mathbf{\!\!}\sum\limits_{j=1}^{K^{(h)}}\sum_{b=1}^{K_{B}}%
\mathbf{\!}\omega _{+}^{(h\mathbf{\!},j,b)\mathbf{\!}}\delta _{\mathbf{\!}%
J^{(h,j)}\cup L^{(b)}\mathbf{\!}}(\mathcal{L(}\mathbf{X}_{+\mathbf{\!}})%
\mathbf{\!})\mathbf{\!\!}\left[ p_{+}^{(h\mathbf{\!})}\right] ^{\mathbf{\!X}%
_{+}}\mathbf{\!},
\end{equation*}%
where%
\begin{eqnarray*}
\omega _{+}^{(h,j,b)} &\triangleq &\omega _{S}^{(I^{(h)},\xi
^{(h)})}(J^{(h,j)})w_{B}(L^{(b)}) \\
p_{+}^{(h)} &\mathbf{\triangleq }&p_{+}^{(\xi ^{(h)})}.
\end{eqnarray*}

Since the weights of the (untruncated) prediction density sum to 1, it
follows from Proposition \ref{Prop_L1_error} that the resulting truncated
density $\mathbf{\hat{\pi}}_{+}=\sum_{h=1}^{H}\mathbf{\hat{\pi}}_{+}^{(h)}$,
which has a total of $T=K_{B}\sum_{h=1}^{H}K^{(h)}$ components, incurs an $%
L_{1}$ truncation error of%
\begin{equation*}
1-\sum_{h=1}^{H}\sum\limits_{j=1}^{K^{(h)}}\sum_{b=1}^{K_{B}}\omega
_{+}^{(h,j,b)}.
\end{equation*}%
Moreover, the truncated density minimizes the $L_{1}$-distance from the
predicted $\delta $-GLMB over all truncations with $K^{(h)}$ components for
each $h=1,...,H$, and $K_{B}$ components for the births. The final
expression for the approximation is obtained by normalizing the truncated
density. Table 2 shows the pseudo code for the prediction operation. Note
that all three for-loops can be implemented in parallel.

\begin{figure}[tbp]
\hrule
\par
\medskip
\par
\textbf{Table 2. Prediction}
\par
\begin{itemize}
\item \textsf{{\footnotesize {input: }}}$\{(I^{(h)},\xi ^{(h)},\omega
^{(h)},p^{(h)},K^{(h)})\}_{h=1}^{H}$%
\par
\item \textsf{{\footnotesize {input: }}}$K_{B}$, $\{(r_{B}^{(\ell
)},p_{B}^{(\ell )})\}_{\ell \in \mathbb{B}}$,
\par
\item \textsf{{\footnotesize {output }}}$\{(I_{+}^{(h,j,b)},\omega
_{+}^{(h,j,b)},p_{+}^{(h)})\}_{(h,j,b)=(1,1,1)}^{(H,K^{(h)},K_{B})}$%
\end{itemize}
\par
\hrule
\par
\medskip
\par
\textsf{{\footnotesize {compute}} }$C_{B}$ \textsf{{\footnotesize {according
to}}} (\ref{eq:costvectorbirth})
\par
$\{L^{(b)}\}_{b=1}^{K_{B}}:=\mathsf{k\_shortest\_path}(\mathbb{B}%
,C_{B},K_{B})$%
\par
\textsf{{\footnotesize {for }}}$b=1:K_{B}$%
\par
\quad $\omega _{B}^{(b)}:=\prod\limits_{\ell \in L^{(b)}}r_{B}^{(\ell
)}\prod\limits_{\ell \in \mathbb{B-}L^{(b)}}\left[ 1-r_{B}^{(\ell )}\right] $%
\par
\textsf{{\footnotesize {end}}}
\par
\textsf{{\footnotesize {for}}}\textsf{\ }$h=1:H$%
\par
\quad $\eta _{S}^{(h)}:=\eta _{S}^{(\xi ^{(h)})}$ \textsf{{\footnotesize {%
according to}}} (\ref{eq:GM_eta_S})/(\ref{eq:Particle_eta_S})
\par
\quad $C^{(h)}:=C^{(I^{(h)},\xi ^{(h)})}$ \textsf{{\footnotesize {according
to}}} (\ref{eq:costKshortest})
\par
\quad $\{J^{(h,j)}\}_{_{j=1}}^{K^{(h)}}:=\mathsf{k\_shortest\_path}%
(I^{(h)},C^{(h)},K^{(h)})$%
\par
\quad \textsf{{\footnotesize {for }}}$(j,b)=(1,1):(K^{(h)},K_{B})$%
\par
\quad \quad $I_{+}^{(h,j,b)}:=J^{(h,j)}\cup L^{(b)}$%
\par
\quad \quad $\omega _{+}^{(h,j,b)}:=\omega ^{(h)}\left[ \eta _{S}^{(h)}%
\right] ^{J^{(h,j)}}\left[ 1-\eta _{S}^{(h)}\right] ^{I^{(h)}-J^{(h,j)}}%
\omega _{B}^{(b)}$%
\par
\quad \textsf{{\footnotesize {end}}}
\par
\quad $p_{+}^{(h)}:=p_{+}^{(\xi ^{(h)})}$ \textsf{{\footnotesize {according
to}}} (\ref{eq:GM_single_predict})/(\ref{eq:Particle_single_predict})
\par
\textsf{{\footnotesize {end}}}
\par
\textsf{{\footnotesize {normalize weights }}}$\{\omega
_{+}^{(h,j,b)}\}_{(h,j,b)=(1,1,1)}^{(H,K^{(h)},K_{B})}$%
\par
\medskip
\par
\hrule
\end{figure}

Specific values for the number of requested components $K^{(h)}$ and $K_{B}$
are generally user specified and application dependent. A generic strategy
is to choose $K^{(h)}=\left\lceil \omega ^{(h)}J_{\max }\right\rceil $ where
$J_{\max }$ is the desired overall number of hypotheses, and to choose $%
K_{B} $ such that the resulting truncation captures a desired proportion
(say 99\%) of the probability mass of the birth density. The alternative
strategy of keeping the $T=J_{\max}$ strongest components of $\mathbf{\pi }%
_{+}$ would yield a smaller $L_{1}$-error than the proposed strategy.
However, in addition to an $(H+K_{B})$-fold increase in the dimension of the
resulting problem, parallelizability is lost.

Remark: As noted previously, the actual value of the association history $%
\xi ^{(h)}$ is not needed in the update and prediction calculations, it is
used merely as an index for the track density $p^{(\xi ^{(h)})}$. Since the
track density is now equivalently indexed by $h$, i.e. $p^{(h)}\triangleq
p^{(\xi ^{(h)})}$, in practice it is not necessary to propagate $\xi ^{(h)}$%
. Nonetheless, for clarity of exposition we have retained $\xi^{(h)}$ in the
update and prediction pseudo codes.

\section{Delta GLMB filter\label{sec:Main}}

The main steps of the $\delta $-GLMB filter algorithm is summarized in the
following pseudo code:

\bigskip 

\hrule

\medskip

\textbf{Main Loop} (Filter)

\medskip

\hrule

\medskip

\textsf{{\footnotesize {for }}}$k=1:K$

\textsf{\footnotesize \qquad Prediction }

\textsf{\footnotesize \qquad Update}

\textsf{\footnotesize \qquad Compute State Estimates}

\textsf{{\footnotesize \textsf{end}}}

\medskip

\hrule

\bigskip 

In Subsection \ref{subsec:stateestimation} we describe the multi-target
state estimation process in the \textquotedblleft Compute State
Estimate\textquotedblright\ module. Subsection \ref{subsec:lookahead}
presents look-ahead strategies to reduce number of calls to the ranked
optimal assignment and K-shortest paths algorithms.

\subsection{Multi-target state estimation\label{subsec:stateestimation}}

Given a multi-target filtering density, several multi-target state
estimators are available. The Joint Multi-object Estimator and Marginal
Multi-object Estimator are Bayes optimal, but difficult to compute \cite%
{Mahler07}. A simple and intuitive multi-target estimator for a $\delta $%
-GLMB density is the multi-Bernoulli estimator, which selects the set of
tracks or labels $L\subseteq $ $\mathbb{L}$ with existence probabilities
above a certain threshold, and for the states of the tracks, the maximum
\emph{a posteriori} (MAP) or the mean estimates from the densities $p^{(\xi
)}(\cdot ,\ell ),\ell \in L$. From \cite{VoConj13}, the existence
probability of track $\ell $ is given by the sum of the weights of all
hypotheses containing track $\ell $ :%
\begin{equation*}
\sum_{(I,\xi )\in \mathcal{F}(\mathbb{L})\times \Xi }\omega ^{(I,\xi
)}1_{I}(\ell ).
\end{equation*}%
An alternative multi-target estimate is the MAP or the mean estimate of the
states of the hypothesis with the highest weight.

In this work we use a suboptimal but tractable version of the Marginal
Multi-object Estimator by finding the MAP cardinality estimate from the
cardinality distribution \cite{VoConj13}:
\begin{equation*}
\mathbf{\rho }(n)=\sum_{(I,\xi )\in \mathcal{F}_{n}(\mathbb{L})\times \Xi
}\omega ^{(I,\xi )},
\end{equation*}%
where $\mathcal{F}_{n}(\mathbb{L})$ denotes the class of finite subsets of $%
\mathbb{L}$ with exactly $n$ elements. We then find the labels and the mean
estimates of the states from the highest weighted component that has the
same cardinality as the MAP cardinality estimate. Table 3 shows the pseudo
code for the multi-target state estimation

\begin{figure}[tbp]
\hrule
\par
\medskip
\par
\textbf{Table 3. Compute State Estimate}
\par
\begin{itemize}
\item {\footnotesize \textsf{input: }}$N_{\max }$, $\{(I^{(h,j)},\xi
^{(h,j)},\omega ^{(h,j)},p^{(h,j)})\}_{(h,j)=(1,1)}^{(H,T^{(h)})}$%
\par
\item {\footnotesize \textsf{output: }}$\mathbf{\hat{X}}$%
\end{itemize}
\par
\medskip
\par
\hrule
\par
\medskip
\par
$\mathbf{\rho }(n):=\sum_{h=1}^{H}\sum_{j=1}^{T^{(h)}}\omega ^{(h,j)}\delta
_{n}(|I^{(h,j)}|)$ ; $n=0,...,N_{\max }$%
\par
$\hat{N}:=\arg \max \mathbf{\rho }$%
\par
$(\hat{h},\hat{\jmath}):=\arg \max_{(h,j)}\omega ^{(h,j)}\delta _{\hat{N}%
}(|I^{(h,j)}|)$%
\par
$\mathbf{\hat{X}:}=\{(x,\ell ):\ell \in I^{(\hat{h},\hat{\jmath})}$, $%
x=\dint yp^{(\hat{h},\hat{\jmath})}(y,\ell )dy\}$%
\par
\medskip
\par
\hrule
\end{figure}

\subsection{PHD look-ahead\label{subsec:lookahead}}

In this subsection we use the PHD/CPHD update and prediction \cite%
{MahlerPHD2, VSD05, VM06, MahlerCPHDAES, VVC07} to look-ahead and identify
the prediction/update components that generate significant updated/predicted
components. This is analogous to the idea of using a measurement driven
proposal in the particle filter, to guide particles towards important areas
of the state space \cite{Doucet2000, Ristic04}.

Recall that each prediction component generates a set of update components.
Typically, more than 90\% of the best predicted components generate update
components with negligible weights. Since updating is expensive, it is
important to minimize the number of prediction components to be updated (and
hence the number of calls to the ranked assignment algorithm). Knowing in
advance which prediction components would generate significant update
components will save substantial computations. Similarly, further saving in
computations can be achieved by knowing in advance which update components
would generate significant prediction components.

The PHD/CPHD filter is a good approximation to the multi-target Bayes filter
and is inexpensive compared to the $\delta $-GLMB update. Moreover,
integration of the PHD filter within the $\delta $-GLMB filter is seamless
as both filters are developed from the same RFS framework. Indeed, the PHD
of (the unlabeled version of) a $\delta $-GLMB of the form (\ref%
{eq:generativeGLMB}) is given by \cite{VoConj13}:
\begin{equation*}
v(x)=\sum_{(I,\xi )\in \mathcal{F}(\mathbb{L})\times \Xi }v^{(I,\xi )}(x)
\end{equation*}%
where%
\begin{equation*}
v^{(I,\xi )}(x)\mathbf{\!}=\sum_{\ell \in I}\omega ^{(I,\xi )}p^{(\xi
)}(x,\ell ).
\end{equation*}%
is the PHD of hypothesis $(I,\xi )$. Assuming that the detection probability
and single measurement likelihood do not depend on the labels, the updated
PHD is given by (see \cite{MahlerPHD2, VSD05, VM06})%
\begin{equation*}
v(x|Z)=\sum_{(I,\xi )\in \mathcal{F}(\mathbb{L})\times \Xi }\hat{v}^{(I,\xi
)}(x|Z),
\end{equation*}%
where the constituent updated PHD $\hat{v}^{(I,\xi )}(\cdot |Z)$\ due to
hypothesis $(I,\xi )$ is given by%
\begin{align*}
\hat{v}^{(I,\xi )}(x|Z)& =(1-p_{D}(x))v^{(I,\xi )}(x) \\
& +\sum_{z\in Z}\frac{p_{D}(x)g(z|x)v^{(I,\xi )}(x)}{\kappa
(z)+\sum\limits_{(I,\xi )\in \mathcal{F}(\mathbb{L})\times \Xi }\left\langle
p_{D}g(z|\cdot ),v^{(I,\xi )}\right\rangle },
\end{align*}%
Note that $\hat{v}^{(I,\xi )}(\cdot |Z)$ is not the same as the updated PHD
of $v^{(I,\xi )}$ given by
\begin{align*}
v^{(I,\xi )}(x|Z)& =(1-p_{D}(x))v^{(I,\xi )}(x) \\
& +\sum_{z\in Z}\frac{p_{D}(x)g(z|x)v^{(I,\xi )}(x)}{\kappa (z)+\left\langle
p_{D}(x)g(z|x),v^{(I,\xi )}(x)\right\rangle }.
\end{align*}

The ratio of the constituent updated PHD mass $\int \hat{v}^{(I,\xi )}(x|Z)dx
$ to the total updated PHD mass $\int v(x|Z)dx$ can be thought of as a
selection criteria and is a good indicator of the significance of hypothesis
$(I,\xi )$ after the update. A higher score indicates a greater
significance. The proposed look-ahead strategy selects prediction hypotheses
with highest constituent updated PHD masses that together makes up most (say
95\%) of the total updated PHD mass. These PHD masses can be readily
computed with $O(\left\vert Z\right\vert )$ complexity, using SMC \cite%
{VSD05} or Gaussian mixtures \cite{VM06}. Further improvement can be
achieved by replacing the PHD update with the CPHD update with $O(\left\vert
Z\right\vert ^{3})$ complexity \cite{MahlerCPHDAES, VVC07}. Note that the $%
\delta $-GLMB update can reuse the computations in the PHD/CPHD update such
as the Kalman gain and other variables.

A similar strategy can be employed to identify updated components that
generates weak prediction hypotheses. Using the constituent predicted PHD
masses as the selection criterion, we select updated hypotheses whose
combined predicted PHD mass makes up most of the total predicted PHD mass.

A parallelizable look-ahead strategy can be formulated using the updated PHD
$v^{(I,\xi )}(\cdot |Z)$. The relative cardinality error $\left\vert \int
v^{(I,\xi )}(x|Z)dx-|I|\right\vert /|I|$ of hypothesis $(I,\xi )$ is a
measure of how well it explains the observed data $Z$, and is a possible
selection criterion. Additional selection criteria are possible with the
CPHD update since the cardinality distribution is available. Consider the
PHD $v^{(I,\xi )}$ of hypothesis $(I,\xi )$ and a Poisson cardinality
distribution $\rho ^{(I,\xi )}$ with mean $\int v^{(I,\xi )}(x)dx$. If
hypothesis $(I,\xi )$ explains the observed data well, then the CPHD updated
cardinality distribution $\rho ^{(I,\xi )}(\cdot |Z)$ should be close to $%
\delta _{|I|}$. Hence, a possible selection criterion is the
Kullback-Leibler divergence between $\rho ^{(I,\xi )}(\cdot |Z)$ and $\delta
_{|I|}$. In both cases a lower score indicates a greater significance for
hypothesis $(I,\xi )$ after the update.

Regardless of the particular look ahead strategy used, the selection
criterion yields an unnormalized score for each hypothesis or component,
which generally indicates how well the component explains the observed data.
Thus for implementation a normalized score for each component can be derived
such that a higher score indicates a better fit. A generic method for
setting $T^{(h)}$ is then to choose its value to be proportional to the
normalized score for component $(I^{(h)},\xi ^{(h)})$ and the desired
overall number of components $J_{\max }$.

\section{Numerical Example\label{sec:Simo}}

A non-linear example has been presented in \cite{VoConj13}. In this section,
we compare the performance of the $\delta $-GLMB and CPHD filters with a
linear Gaussian example and hence Gaussian mixture implementation. A typical
scenario is employed to highlight the susceptibility of the CPHD filter to
the so called \textquotedblleft spooky effect\textquotedblright\ \cite%
{Franken09}. The term spooky effect refers to phenomenon where the CPHD
filter encounters a missed detection for a particular track and
significantly reduces the weight of the undetected track by shifting part of
its mass to other targets. The $\delta $-GLMB filter is generally immune to
the spooky effect since it does not use an i.i.d. approximation of the
filtering density and hence cannot shift the probability mass of individual
tracks between each other. Furthermore the $\delta $-GLMB filter is
generally able to correct for CPHD look ahead errors, since the weights of the CPHD-generated hypotheses are computed exactly from the $\delta $-GLMB update.

\begin{figure}[tbh]
\begin{center}
\resizebox{80mm}{!}{\includegraphics[clip=true]{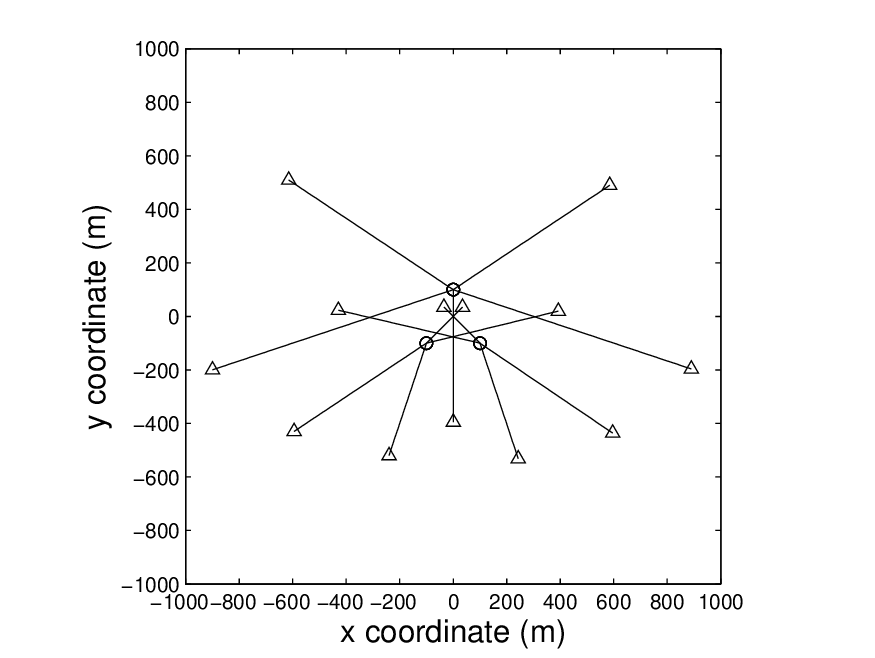}}
\end{center}
\caption{Multiple trajectories in the $xy$ plane. Start/Stop positions for
each track are shown with {\protect\large $\circ $}/{\protect\small $%
\triangle $}. Targets become more dispersed with increasing time.}
\label{fig:xy}
\end{figure}

Consider a set of multi-target trajectories on the two dimensional region $%
[-1000,1000]m\times \lbrack -1000,1000]m$, as shown in Figure \ref{fig:xy}. The duration of the scenario is $K=100s$. All targets travel in straight paths and with different but
constant velocities. The number of targets is time varying due to births and
deaths. There is a crossing of 3 targets at the origin at time $k=20$, and a
crossing of two pairs of targets at position $(\pm 300,0)$ at time $k=40$.
The targets also become more dispersed as the time increases in order to
elucidate the estimation errors caused by the spooky effect.

The kinematic target state is a vector of planar position and velocity $%
x_{k}=[~p_{x,k},p_{y,k},\dot{p}_{x,k},\dot{p}_{y,k}~]^{T}$. Measurements are
noisy vectors of planar position only $z_{k}=[~z_{x,k},z_{y,k}~]^{T}$. The
single-target state space model is linear Gaussian according to transition
density $f_{k|k-1}(x_{k}|x_{k-1})=\mathcal{N}(x_{k};F_{k}x_{k-1},Q_{k})$ and
likelihood $g_{k}(z_{k}|x_{k})=\mathcal{N}(z_{k};H_{k}x_{k},R_{k})$ with
parameters%
\begin{equation*}
\begin{array}{cc}
F_{k}=%
\begin{bmatrix}
I_{2} & \Delta I_{2} \\
0_{2} & I_{2}%
\end{bmatrix}
& Q_{k}=\sigma _{\nu }^{2}%
\begin{bmatrix}
\frac{\Delta ^{4}}{4}I_{2} & \frac{\Delta ^{3}}{2}I_{2} \\
\frac{\Delta ^{3}}{2}I_{2} & \Delta ^{2}I_{2}%
\end{bmatrix}
\\
H_{k}=%
\begin{bmatrix}
I_{2} & 0_{2}%
\end{bmatrix}
& R_{k}=\sigma _{\varepsilon }^{2}I_{2}%
\end{array}%
\end{equation*}%
where $I_{n}$ and $0_{n}$ denote the $n\times n$ identity and zero matrices
respectively, $\Delta =1s$ is the sampling period, $\sigma _{\nu }=5m/s^{2}$
and $\sigma _{\varepsilon }=10m$ are the standard deviations of the process
noise and measurement noise. The survival probability is $p_{S,k}=0.99$ and
the birth model is a Labeled Multi-Bernoulli RFS with parameters $\pi
_{B}=\{r_{B}^{(i)},p_{B}^{(i)}\}_{i=1}^{3}$ where $r_{B}^{(i)}=0.04$ and $%
p_{B}^{(i)}(x)=\mathcal{N}(x;m_{B}^{(i)},P_{B})$ with $%
m_{B}^{(1)}=[~0,0,100,0~]^{T}$, $m_{B}^{(2)}=[~-100,0,-100,0~]^{T}$, $%
m_{B}^{(3)}=[~100,0,-100,0~]^{T}$, $P_{B}=\mathrm{diag}%
([~10,~10,~10,~10~]^{T})^{2}$. The detection probability is $p_{D,k}=0.88$\
and clutter follows a Poisson RFS with an average intensity of $\lambda
_{c}=1.65\times 10^{-5}~m^{-2}$ giving an average of 66 false alarms per
scan.

\begin{figure}[h]
\begin{center}
\resizebox{80mm}{!}{\includegraphics{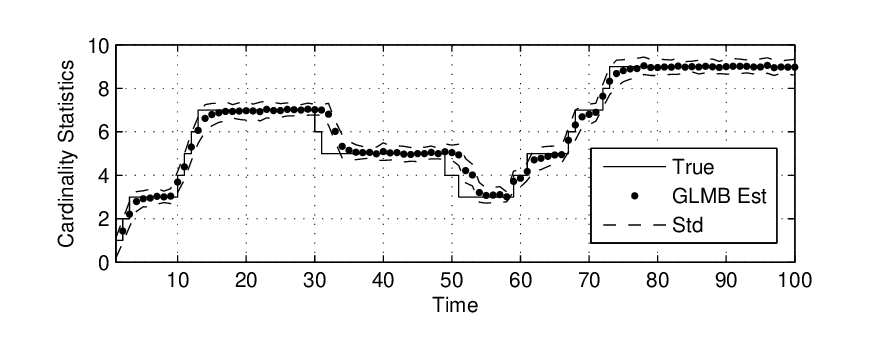}}
\end{center}
\caption{Cardinality statistics for $\protect\delta $-GLMB filter (100 Monte Carlo trials)}
\label{fig:cdnplot1}
\end{figure}

\begin{figure}[h]
\begin{center}
\resizebox{80mm}{!}{\includegraphics{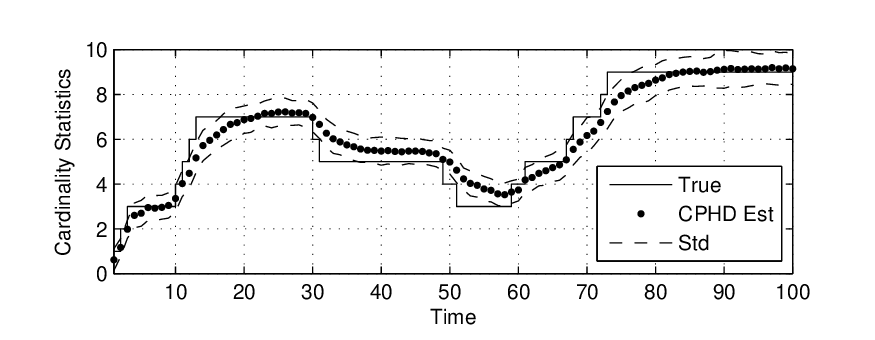}}
\end{center}
\caption{Cardinality statistics for CPHD filter (100 Monte Carlo trials)}
\label{fig:cdnplot2}
\end{figure}

\begin{figure}[h]
\begin{center}
\resizebox{80mm}{!}{\includegraphics{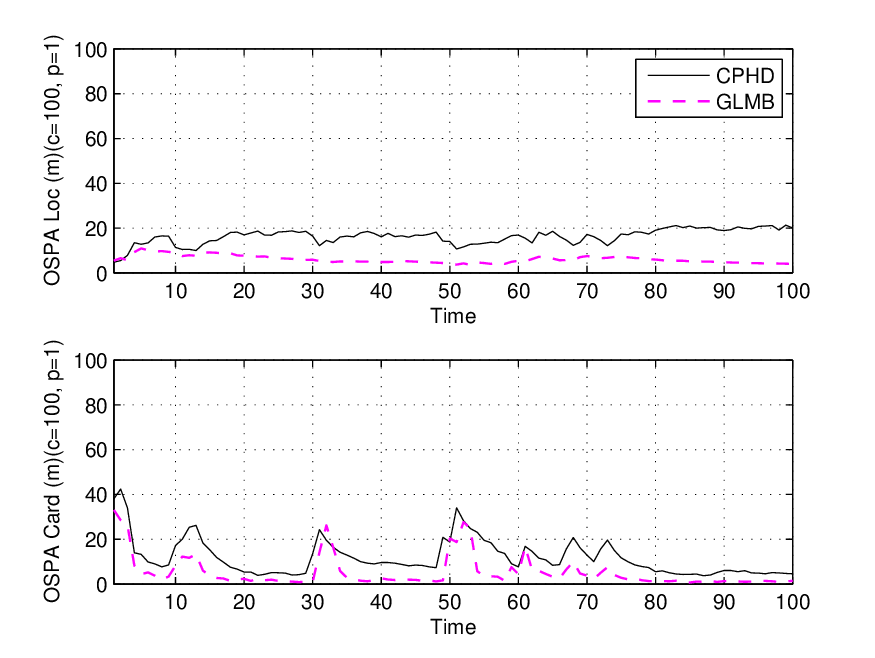}}
\end{center}
\caption{OSPA components for $\protect\delta $-GLMB and CPHD filters (100 Monte Carlo trials)
}
\label{fig:ospacomp}
\end{figure}

The $\delta $-GLMB filter is capped to 10000 components and is coupled with
the parallel CPHD look ahead strategy described in the previous section. The
CPHD filter is similarly capped to 10000 components through pruning and
merging of mixture components. Results are shown over 100 Monte Carlo
trials. Figures \ref{fig:cdnplot1} and \ref{fig:cdnplot2} show the mean and
standard deviation of the estimated cardinality versus time. Figures \ref%
{fig:ospatot} and \ref{fig:ospacomp} show the Optimal Sub-Pattern Assignment
(OSPA)\ distance \cite{Schumacher08} and its localization and cardinality
components for $c=100m$ and $p=1$.

It can be seen that both filters estimate the target cardinality accurately,
with the $\delta $-GLMB exhibiting better estimated cardinality variance.
However the $\delta $-GLMB filter significantly outperforms the CPHD filter
on the overall miss distance. Examination of the localization and
cardinality components reveals that the $\delta $-GLMB filter outperforms
the CPHD filter on both components. The improved cardinality performance is
attributed mainly due to a lower estimated cardinality variance. The
improved localization performance is attributed to two factors: (a) the
spooky effect causes CPHD filter to temporarily drop tracks which are
subjected to missed detections and to declare multiple estimates for
existing tracks in place of the dropped tracks, and (b) the $\delta $-GLMB
filter is generally able to better localize targets due to a more accurate
propagation of the filtering density.

\begin{figure}[h]
\begin{center}
\resizebox{80mm}{!}{\includegraphics{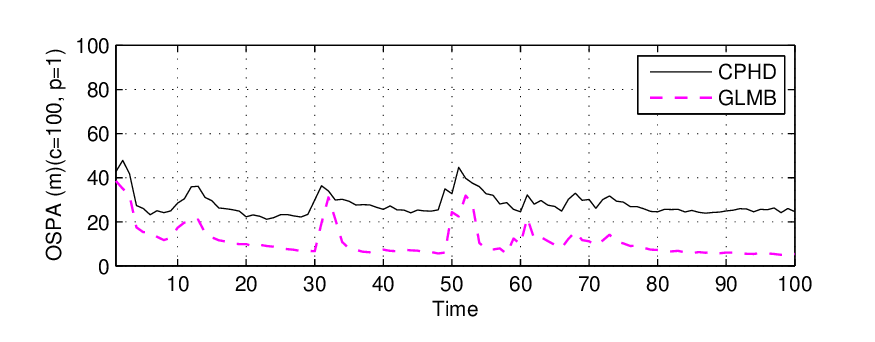}}
\end{center}
\caption{OSPA distance for $\protect\delta $-GLMB and CPHD filters (100 Monte Carlo trials)
}
\label{fig:ospatot}
\end{figure}

\section{Concluding Remarks\label{sec:Conclusion}}

This paper detailed the first implementation of the $\delta $-GLMB
multi-target tracking filter that is general enough to accommodate unknown
and time-varying number of targets, non-linear target dynamics, non-uniform
probability of detection and clutter intensity. A salient feature of this
implementation is the high parallelizability. The key innovation lies in the
truncation of $\delta $-GLMB densities without exhaustive computation of all
the components and the integration of PHD look-ahead to reduce the number of
computations. Furthermore, it is established that truncation of a $\delta $%
-GLMB density by keeping the highest weighted components minimizes the $%
L_{1} $-error in the multi-target densities.

It is also of interest to examine information theoretic criteria, such as
the Kullback-Leibler divergence, for $\delta $-GLMB
truncation/approximation. Implementations using other multi-target filters
(in place of the ranked assignment algorithm) to generate the significant $%
\delta $-GLMB components can be explored to reduce numerical complexity.
Approximation of $\delta $-GLMB by other families, for example the labeled
multi-Bernoulli--a GLMB with only one term \cite{Reuter14}--is another venue
for further work.


\begin{thebibliography}{99}
\bibitem{VoConj13} B.-T. Vo, and B.-N. Vo, \textquotedblleft Labeled Random
Finite Sets and multi-object conjugate priors,\textquotedblright\ \emph{IEEE
Trans. Signal Processing}, Vol. 61, No. 13, pp. 3460--3475, 2013.

\bibitem{Blackman} S. Blackman and R. Popoli, \emph{Design and Analysis of
Modern Tracking Systems}, Artech House, 1999.

\bibitem{Mahler07} R. Mahler, \emph{Statistical Multisource-Multitarget
Information Fusion}, Artech House, 2007.

\bibitem{Bar88} Y. Bar-Shalom, P. K. Willett, and X. Tian, \emph{Tracking
and Data Fusion: A Handbook of Algorithms}, YBS Publishing, 2011.

\bibitem{Reid77} D.~Reid, \textquotedblleft An algorithm for tracking
multiple targets,\textquotedblright\ \emph{IEEE Trans. Automatic Control},
Vol. 24, No.~6, pp. 843--854, 1979.

\bibitem{Kurien90} T. Kurien, \textquotedblleft Issues in the design of
Practical Multitarget Tracking algorithms,\textquotedblright\ in \emph{%
Multitarget-Multisensor Tracking: Advanced Applications}, Y. Bar-Shalom
(Ed), Artech House, pp. 43--83, 1990.

\bibitem{Mallick12} M. Mallick, S. Coraluppi, and C. Carthel,
\textquotedblleft Multi-target tracking using Multiple Hypothesis
Tracking,\textquotedblright\ in \emph{Integrated Tracking, Classification,
and Sensor Management: Theory and Applications}, M. Mallick, V.
Krishnamurthy, B.-N. Vo (Eds.), Wiley/IEEE, pp. 165--201, 2012.

\bibitem{MahlerPHD2} R. Mahler, \textquotedblleft Multi-target Bayes
filtering via first-order multi-target moments,\textquotedblright\ \emph{%
IEEE Transactions of Aerospace and Electronic Systems}, Vol. 39, No. 4, pp.
1152--1178, 2003.

\bibitem{VVPS10} B.-N. Vo, B.-T. Vo, N.-T. Pham, and D. Suter,
\textquotedblleft Joint detection and estimation of multiple objects from
image observations,\textquotedblright\ \emph{IEEE Trans. Signal Procesing},
Vol. 58, No. 10, pp. 5129--5241, 2010.

\bibitem{Clark07} D. E. Clark, I. T. Ruiz, Y. Petillot, and J. Bell,
\textquotedblleft Particle PHD filter multiple target tracking in sonar
image,\textquotedblright\ \emph{IEEE Trans. Aerospace \& Electronic Systems}%
, Vol. 43, No. 1, pp. 409--416, 2007.

\bibitem{Pham07} N. T. Pham, W. Huang, and S. H. Ong, \textquotedblleft
Tracking multiple objects using Probability Hypothesis Density filter and
color measurements,\textquotedblright \emph{\ IEEE Int. Conf. Multimedia and
Expo}, pp. 1511--1514, July 2007,.

\bibitem{Maggio08} E. Maggio, M. Taj, and A. Cavallaro,\textquotedblleft
Efficient multitarget visual tracking using random finite
sets,\textquotedblright\ \emph{IEEE Trans. Circuits \& Systems for Video
Technology}, Vol. 18, No. 8, pp. 1016--1027, 2008.

\bibitem{Hosy12} R. Hoseinnezhad, B.-N.Vo, D. Suter, and B.-T. Vo,
\textquotedblleft Visual tracking of numerous targets via multi-Bernoulli
filtering of image data,\textquotedblright\ \emph{Pattern Recognition}, Vol.
45, No. 10, pp. 3625--3635, 2012.

\bibitem{Hosy13} R. Hoseinnezhad, B.-N. Vo and B.-T. Vo, \textquotedblleft
Visual tracking in background subtracted image sequences via multi-Bernoulli
filtering\textquotedblright , \emph{IEEE Trans. Signal Processing}, Vol. 61,
No. 2, pp. 392--397, 2013.

\bibitem{Mullane11} J. Mullane, B.-N. Vo, M. Adams, and B.-T. Vo,
\textquotedblleft A Random Finite Set approach to Bayesian
SLAM,\textquotedblright\ \emph{IEEE Trans. Robotics}, Vol. 27, No. 2, pp.
268--282, 2011.

\bibitem{Zhang12} F. Zhang, H. Stahle, A. Gaschler, C. Buckl, and A. Knoll,
\textquotedblleft Single camera visual odometry based on Random Finite Set
Statistics,\textquotedblright\ \emph{IEEE/RSJ Int. Conf. Intelligent Robots
and Systems (IROS)}, pp. 559--566, Oct. 2012.

\bibitem{Morat12} D. Moratuwage, B.-N. Vo, D. Wang, and H. Wang,
\textquotedblleft Extending Bayesian RFS SLAM to multi-vehicle
SLAM\textquotedblright , \emph{12th Int. Conf. Control, Automation, Robotics
\& Vision}, (ICARCV'12), Guangzhou, China, Dec. 2012.

\bibitem{Morat13} D. Moratuwage, D. Wang, and B.-N. Vo, "Collaborative
multi-vehicle SLAM with moving object tracking\textquotedblright , \emph{%
IEEE Int. Conf. Robotics \& Automation}, (ICRA'13), Karlsruhe, Germany, May
2013.

\bibitem{Lee13} C.-S. Lee, D.E. Clark, and J. Salvi, \textquotedblleft SLAM
with dynamic targets via single-cluster PHD filtering,\textquotedblright\
\emph{IEEE Journal of Selected Topics in Signal Processing}, Vol. 7, No. 3,
pp. 543--552, 2013.

\bibitem{Battis08} G. Battistelli, L. Chisci, S. Morrocchi, F. Papi, A.
Benavoli, A. D. Lallo, A. Farina, and A. Graziano, \textquotedblleft Traffic
intensity estimation via PHD filtering,\textquotedblright\ \emph{5th
European Radar Conf.}, Amsterdam, The Netherlands, pp. 340--343, Oct. 2008.

\bibitem{Mila13} L. Mihaylova, \textquotedblleft Probability Hypothesis
Density filtering for real-time traffic state estimation and
prediction,\textquotedblright\ \emph{Network and Heterogeneous Media. An
Applied Mathematics Journal}, 2013.

\bibitem{Meissner13} D. Meissner, S. Reuter, and K. Dietmayer,
\textquotedblleft Road user tracking at intersections using a multiple-model
PHD filter,\textquotedblright\ \emph{IEEE Intelligent Vehicles Symposium}
(IV), pp. 377--382, June 2013.

\bibitem{Juang09} R. Juang, A. Levchenko, and P. Burlina, \textquotedblleft
Tracking cell motion using GM-PHD,\textquotedblright\ \emph{Int. Symp.
Biomedical Imaging}, pp. 1154--1157, June 2009.

\bibitem{Hamid13} S. H. Rezatofighi, S. Gould, B.-N. Vo, K. Mele, W. Hughes,
and R. Hartley, \textquotedblleft A multiple model Probability Hypothesis
Density tracker for time-lapse cell microscopy sequences,\textquotedblright
\emph{\ Int. Conf. Inf. Proc. Medical Imaging} (IPMI), May 2013.

\bibitem{Zhang11} X. Zhang, \textquotedblleft Adaptive control and
reconfiguration of mobile wireless sensor networks for dynamic multi-target
tracking,\textquotedblright\ \emph{IEEE Trans. Automatic Control,} Vol. 56,
No. 10, pp. 2429--2444, 2011.

\bibitem{LeeYao12} J. Lee and K. Yao, \textquotedblleft Initialization of
multi-Bernoulli Random Finite Sets over a sensor tree,\textquotedblright\
\emph{Int. Conf. Acoustics, Speech \& Signal Processing} (ICASSP), pp.
25--30, Mar. 2012.

\bibitem{Battis13} G. Battistelli, L. Chisci, C. Fantacci, A. Farina, and A.
Graziano, \textquotedblleft Consensus CPHD filter for distributed
multitarget tracking,\textquotedblright\ \emph{IEEE Journal on Selected
Topics in Signal Processing}, Vol. 7, No. 3, pp. 508--520, 2013.

\bibitem{Uney13} M. Uney, D.E Clark, S.J. Julier, \textquotedblleft
Distributed fusion of PHD filters via exponential mixture
densities,\textquotedblright\ \emph{IEEE Journal on Selected Topics in
Signal Processing}, Vol. 7, No. 3, pp. 521--531, 2013.

\bibitem{MahlerCPHDAES} R.~Mahler, \textquotedblleft {P}{H}{D} filters of
higher order in target number,\textquotedblright\ \emph{IEEE Trans.
Aerospace \& Electronic Systems}, Vol. 43, No. 3, pp. 1523--1543 2007.

\bibitem{VVC09} B.-T. Vo, B.-N. Vo, and A. Cantoni, \textquotedblleft The
cardinality balanced multi-target multi-Bernoulli filter and its
implementations,\textquotedblright\ \emph{IEEE Trans. Signal Processing}, \
Vol. 57, No. 2, pp. 409--423, 2009.

\bibitem{VoVo11} B.-T. Vo, and B.-N. Vo, \textquotedblleft A Random Finite
Set conjugate prior and application to multi-target
tracking,\textquotedblright\ Proc. 7th Int. \emph{Conf. Intelligent Sensors,
Sensor Networks \& Information Processing} (ISSNIP'2011), Adelaide,
Australia, Dec. 2011.

\bibitem{Daley88} D.J. Daley and D. Vere-Jones, \emph{An Introduction to the
Theory of Point Processes}, Springer, New York, 1988.

\bibitem{Stoyanetal} D. Stoyan, W., Kendall, and J. Mecke
\emph{Stochastic geometry and its applications} (2nd Ed). Chichester: Wiley,
1995.

\bibitem{Vothesis08} B.-T. Vo, \emph{Random Finite Sets in Multi-Object
Filtering}, PhD Thesis, University of Western Australia, 2008.

\bibitem{Ristic13} B. Ristic, \emph{Particle Filters for Random Set Models},
Springer, 2013.

\bibitem{Kuhn55} H. Kuhn, \textquotedblleft The Hungarian method for the
assignment problem,\textquotedblright\ \emph{Naval Research Logistics
Quarterly}, Vol. 2, pp. 83--97, 1955

\bibitem{Murty} K. G. Murty, \textquotedblleft An algorithm for ranking all
the assignments in order of increasing cost,\textquotedblright\ \emph{%
Operations Research}, Vol. 16, No. 3, pp. 682--687, 1968.

\bibitem{Munkres57} J. Munkres, \textquotedblleft Algorithms for assignment
and transportation problems,\textquotedblright\ \emph{Journal of the Society
for Industrial and Applied Mathematics}, Vol. 5, No. 1, Mar. 1957.

\bibitem{Jonker87} R. Jonker, and T. Volgenant, \textquotedblleft A shortest
augmenting path algorithm for dense and sparse linear assignment
problems,\textquotedblright\ \emph{Computing} Vol. 38, No. 11, pp. 325--340,
Nov. 1987.

\bibitem{DanchickNewnam93} R. Danchick and G. E. Newman, \textquotedblleft A
fast method for finding the exact N-best hypotheses for multitarget
tracking\textquotedblright , \emph{IEEE Trans. Aerospace \& Electronic
Systems}, Vol. 29, No. 2, pp. 555--560, 1993.

\bibitem{CoxMiller95} I. Cox, and M. Miller, \textquotedblleft On finding
ranked assignments with application to multitarget tracking and motion
correspondence,\textquotedblright\ \emph{IEEE Trans. Aerospace \& Electronic
Systems}, Vol. 32, No. 1, pp. 486--489, 1995.

\bibitem{CoxHingorani96} I. Cox, and S. Hingorani \textquotedblleft An
efficient implementation of Reid's multiple hypothesis tracking algorithm
and its evaluation for the purpose of visual tracking,\textquotedblright\
\emph{IEEE Trans. Pattern Analysis \& Machine Intelligence}, Vol. 18, No. 2,
pp. 138--150, 1996.

\bibitem{Pattipati} K. R. Pattipati, R. L. Popp, and T. T. Kirubarajan,
\textquotedblleft Survey of Assignment Techniques for Multitarget
Tracking,\textquotedblright\ in \emph{Multitarget-Multisensor Tracking:
Applications and Advances}, Vol. 3, Y. Bar-Shalom and D. Blair (Eds.),
Artech House, pp. 77--159, 2000.

\bibitem{Milleretal97} M. Miller, H. Stone, and I. Cox, \textquotedblleft
Optimizing Murty's ranked assignment method,\textquotedblright\ \emph{IEEE
Trans. Aerospace \& Electronic Systems}, Vol. 33, No. 3, pp. 851--862, 1997.

\bibitem{Pascoaletal03} M. Pascoal, M. Captivo, and J. Cl\i maco,
\textquotedblleft A note on a new variant of Murty's ranking assignments
algorithm,\textquotedblright\ \emph{4OR: Quarterly Journal of the Belgian,
French and Italian Operations Research Societies}, Vol. 1, No. 3, pp.
243--255, 2003.

\bibitem{Pedersenetal08} C. Pedersen, L. Nielsen, and K. Andersen,
\textquotedblleft An algorithm for ranking assignments using reoptimization,
\textquotedblright\ \emph{Computers \& Operations Research}, Vol. 35, No.
11, pp. 3714--3726, 2008.

\bibitem{Eppstein98finding} D. Eppstein, \textquotedblleft Finding the k
shortest paths,\textquotedblright\ \emph{SIAM Journal on computing}, Vol.
28, No. 2, pp. 652--673, 1998.

\bibitem{Bellman52routing} R. Bellman, \textquotedblleft On a routing
problem,\textquotedblright\ \emph{Quarterly of Applied Mathematics}, Vol.
16, pp. 87--90, 1958

\bibitem{FordFulkersonbk62} L. R. Ford, and D. R. Fulkerson, \emph{Flow in
Networks}, Princeton University Press, 1962.

\bibitem{VSD05} B.-N. Vo, S. Singh and A. Doucet, \textquotedblleft
Sequential Monte Carlo methods for multi-target filtering with Random Finite
Sets,\textquotedblright\ \emph{IEEE Trans.} \emph{Aerospace \& Electronic
Systems}, Vol. 41, No. 4, pp. 1224--1245, 2005.

\bibitem{VM06} B.-N. Vo and W.-K. Ma, \textquotedblleft The Gaussian mixture
Probability Hypothesis Density filter\textquotedblright ,\ \emph{IEEE Trans.
Signal Processing}, Vol. 54, No. 11, pp. 4091--4104, 2006.

\bibitem{VVC07} B.-T. Vo, B.-N. Vo, and A. Cantoni, \textquotedblleft
Analytic implementations of the Cardinalized Probability Hypothesis Density
filter,\textquotedblright\ \emph{IEEE Trans. Signal Processing}, \ Vol. 55,
No. 7, pp. 3553--3567, 2007.

\bibitem{Doucet2000} A. Doucet, S. J. Godsill, and C. Andrieu,
\textquotedblleft On sequential Monte Carlo sampling methods for Bayesian
filtering,\textquotedblright\ \emph{Stat. Comp.}, Vol. 10, pp. 197--208,
2000.

\bibitem{Ristic04} B. Ristic, S. Arulampalam, and N. Gordon, \emph{Beyond
the Kalman filter: Particle Filters for Tracking Applications}, Artech
House, 2004.

\bibitem{Franken09} D. Franken, M. Schmidt, and M. Ulmke, \textquotedblleft
Spooky action at a distance in the Cardinalized Probability Hypothesis
Density filter,\textquotedblright\ \emph{IEEE Trans. Aerospace and
Electronic Systems}, Vol.~45, No.~4, pp. 1657--1664, 2009.

\bibitem{Schumacher08} D. Schuhmacher, B.-T. Vo, and B.-N. Vo,
\textquotedblleft A consistent metric for performance evaluation of
multi-object filters,\textquotedblright\ \emph{IEEE Trans. Signal Processing}%
, Vol. 56, No. 8, pp. 3447--3457, 2008.

\bibitem{Reuter14} S. Reuter, B.-T. Vo, B.-N. Vo, and K. Dietmayer, "The
labelled multi-Bernoulli filter," \emph{IEEE Trans. Signal Processing}, Vol.
62, No. 12, pp. 3246--3260, 2014.
\end{thebibliography}
\end{document}